%% file: draft.tex
\documentclass[reprint,twocolumn,superscriptaddress,showpacs,amsmath,amssymb,aps,prl]{revtex4-1}

\usepackage{graphicx,epstopdf}
\usepackage{subfigure}
\usepackage[usenames,dvipsnames]{color}
\usepackage{slashed}
\usepackage{hyperref}
\usepackage[utf8]{inputenc}
\def\parsla{\partial\!\!\!\slash}
\makeatother

\newcommand{\GeV}{{\rm GeV}}

\allowdisplaybreaks[2]

\begin{document}
\title{Exploring for sub-MeV Boosted Dark Matter from Xenon Electron Direct Detection}

\author{Qing-Hong Cao}
\email{qinghongcao@pku.edu.cn}
\affiliation{Center for High Energy Physics, Peking University, Beijing 100871, China}
\affiliation{Department of Physics and State Key Laboratory of Nuclear Physics and Technology, Peking University, Beijing 100871, China}

\author{Ran Ding}
\email{dingran@mail.nankai.edu.cn}
\affiliation{School of Physics and Materials Science, Anhui University, Hefei 230039, China}
\affiliation{Center for High Energy Physics, Peking University, Beijing 100871, China}

\author{Qian-Fei Xiang}
\email{xiangqf@pku.edu.cn}
\affiliation{Center for High Energy Physics, Peking University, Beijing 100871, China}

\begin{abstract}
\noindent
Direct detection experiments turn to lose sensitivity of searching for a sub-MeV light dark matter candidate
due to the threshold of recoil energy. However, such light dark matter particles can be accelerated by energetic cosmic-rays such that they can be detected with existing detectors. We derive the constraints on the scattering of a boosted light dark matter and electron from the XENON100/1T experiment. We illustrate that the energy dependence of the cross section plays a crucial role in improving both the detection sensitivity and also the complementarity of direct detection and other experiments.
\end{abstract}
\maketitle

Light dark matter (DM) candidate is well motivated and can be naturally realized when the  DM candidate couples feebly to visible sector~\cite{Hall:2009bx,Chu:2011be,Essig:2011nj,Knapen:2017xzo,Bernal:2017kxu}. In particular, it is difficult for a sub-MeV DM candidate to satisfy observed relic abundance through the thermal freeze-out mechanism~\cite{Boehm:2013jpa,Nollett:2013pwa,Cao:2018nbr}; therefore, freeze-in via annihilation of electron-positron pairs is a primary mechanism for DM production ~\cite{Chu:2011be,Essig:2011nj,Dvorkin:2019zdi}. The traditional direct detection of DM-nucleus scattering loses sensitivity rapidly for a DM candidate whose mass is below $\sim{\rm GeV}$ due to the threshold of recoil energy. An alternative way to search for a light DM candidate is through the scattering off electrons~\cite{Essig:2011nj,Essig:2012yx,Essig:2017kqs}, which is not sensitive to a sub-MeV DM candidate neither. It is crucial to develop new approach to probe freeze-in DM in such mass range.

A certain fraction of DM candidates in the Galactic halo would be accelerated by energetic Cosmic-Ray (CR) particles as long as the DM candidate interacts with SM particles. The CR-boosted mechanism relaxes the threshold problem and improves the sensitivity of detecting a light DM candidate~\cite{An:2017ojc,Bringmann:2018cvk}. It has been extensively discussed in DM-nucleus direct detections, neutrino experiments and CR observations for various DM models~\cite{Cappiello:2018hsu,Ema:2018bih,Alvey:2019zaa,Cappiello:2019qsw,Dent:2019krz,Krnjaic:2019dzc,Bondarenko:2019vrb,Berger:2019ttc,Wang:2019jtk}.
In this Letter we investigate the CR-boosted effect on the DM-electron direct detection in the freeze-in scenario and show that the existing data from xenon experiments are able to probe a sub-MeV DM candidate.

For illustration, we consider a typical freeze-in DM model based on the vector-portal, in which the DM candidate is a Dirac fermion ($\chi$) that couples to the visible sector through an additional gauge boson $A^\prime_\mu$,
named as ``dark photon".  The Lagrangian is given by
\begin{equation}
\mathcal{L} \supset \overline{\chi}(i\parsla-m_{\chi})\chi + g_\chi \overline{\chi} \gamma^\mu \chi A^\prime_\mu + g_{\rm SM}\overline{e} \gamma^{\mu} e  A^\prime_\mu +\frac{1}{2} m^2_{A^\prime}A^\prime_\mu A^{\prime \mu}\,,
\label{eq:lag}
\end{equation}
where $m_{\chi}$ and $m_{A^\prime}$ denote the mass of DM candidate and the dark photon, respectively.  $g_\chi$ and $g_{\rm SM}$ are the coupling strength of $A^\prime$ to the DM candidate and the electron, respectively. When the DM candidate scatters off an incident CR electron with a given kinetic energy ($T_{\rm CR}$), the distribution of the DM recoil energy $T_\chi$ is
\begin{eqnarray}
&&\frac{d \sigma_{\chi e}}{d T_{\chi}}
\label{eq:dsdT}
 =\bar{\sigma}_e \frac{\left(\alpha^2 m_e^2+m_{A^\prime}^{2}\right)^{2}}{\mu_{\chi e}^{2}} \times \\\nonumber
&&\frac{2 m_{\chi}\left(m_e+T_{\rm CR}\right)^{2}-T_{\chi}\left(\left(m_e+m_{\chi}\right)^{2}+2 m_{\chi} T_{\rm CR}\right)+m_{\chi} T_{\chi}^{2}}{4 \left(2 m_e T_{\rm CR}+T_{\rm CR}^{2}\right)\left(2 m_{\chi} T_{\chi}+m_{A^\prime}^{2}\right)^{2}}\,,
\end{eqnarray}
where $\bar{\sigma}_e$ denotes the cross section of  DM-free electron scattering for a fixed momentum transfer $q=\alpha m_e$~\cite{Essig:2011nj}.
The maximal recoil energy of the DM candidate is~\cite{goldstein:2001}
\begin{equation}
T_\chi^{\rm max}=\frac{2 m_\chi T_{\rm CR}(T_{\rm CR}+2m_e)}{(m_e+m_\chi)^2+2T_{\rm CR}m_\chi}\,.
\end{equation}
Convoluting the $T_\chi$ distribution in Eq.~(\ref{eq:dsdT}) with the energy spectrum of incident CR electrons $d\Phi_e/dT_{\rm CR}$ yields the recoil flux of boosted DM candidate~\cite{Bondarenko:2019vrb} \begin{equation}
\frac{d\Phi_\chi}{dT_\chi}= D_{\rm eff}\frac{\rho^{\rm local}_\chi}{m_\chi}\int_{T_{\rm CR}^{\rm min}}^{\infty}dT_{\rm CR}\frac{d\Phi_e}{dT_{\rm CR}}\frac{d\sigma_{\chi e}}{dT_\chi}\,,
\label{eq:DMrecoil}
\end{equation}
where $D_{\rm eff}\equiv \int \frac{d\Omega}{4\pi}\int_{l.o.s}dl$ is an effective diffusion distance. See supplement materials for details. For a homogeneous CR distribution and NFW DM halo profile~\cite{Navarro:1995iw,Navarro:1996gj} (scale radius $r_s =20$ kpc and local DM density $\rho^{\rm local}_\chi=0.4\,\GeV\,{\rm cm}^{-3}$), integrating along the line-of-sight to 10 kpc yields $D_{\rm eff}=8.02~{\rm kpc}$~\cite{Bringmann:2018cvk}. In order to produce a recoil energy $T_\chi$ after the DM and CR-electron scattering, the minimum kinetic energy ($T_{\rm CR}^{\rm min}$) of the incident CR electron is given by
\begin{equation}
T_{\rm CR}^{\rm min}=\left(\frac{T_\chi}{2}-m_e\right)\left(1 \pm \sqrt{1+\frac{2T_\chi}{m_\chi}\frac{(m_e+m_\chi)^2}{(2m_e-T_\chi)^2}}\right)\,,
\label{eq:Tmin}
\end{equation}
where the plus and minus sign corresponds to $T_\chi > 2m_e$ and $T_\chi <2m_e$, respectively.

\begin{figure}
\centering
\includegraphics[width=0.45\textwidth]{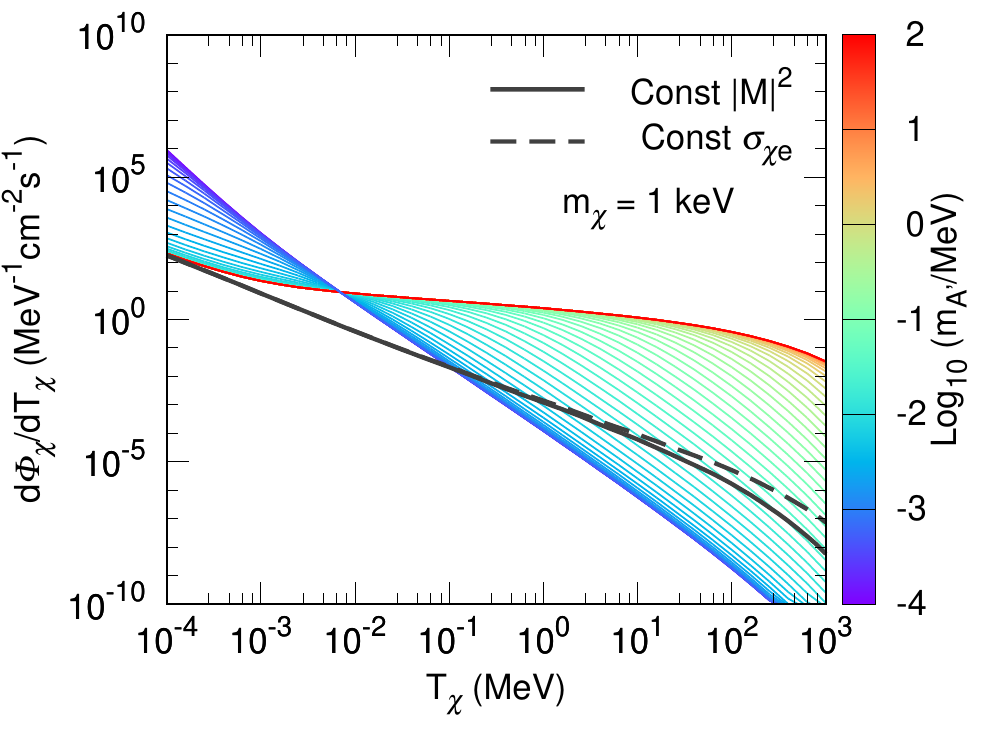}
\caption{Recoil flux distributions of the DM candidate for varying  for $m_{A^\prime}$'s with the choice of $m_\chi=1$ keV and $\bar{\sigma}_e=10^{-30}~{\rm cm}^2$. For comparison, the recoil flux distributions for the approximation of a constant $\sigma_{\chi e}$ (black solid) and a constant $\overline{|\mathcal{M}|^{2}}$ (black dashed) are also plotted. }
\label{fig:DMflux}
\end{figure}

Figure~\ref{fig:DMflux} plots the recoil flux $d\Phi_\chi/dT_\chi$ distributions as a function of $T_\chi$ for various $m_{A^\prime}$'s. Two simplified models are also plotted for comparison; one is the cross section $\sigma_{\chi e}$ being a constant (black-solid curve), the other is that the
the squared matrix element of the DM-electron scattering ($\overline{|\mathcal{M}|^{2}}$), averaged over initial and summed over final spin states,  is a constant (black-dashed curve), i.e.
\begin{equation}
\frac{d\sigma_{\chi e}}{dT_\chi}=
\begin{cases}\dfrac{\bar{\sigma}_{e}}{T^{\max}_\chi} , &
\sigma_{\chi e}={\rm const},\\
\dfrac{\bar{\sigma}_{e}}{T^{\max}_\chi}\dfrac{\left(m_{\chi}+m_{e}\right)^{2}}{\left(m_{\chi}+m_{e}\right)^{2}+2 m_{\chi} T_{\rm CR}}, &  \overline{|\mathcal{M}|^{2}}={\rm const}.
\end{cases}
\end{equation}
The former case is commonly used in the study of non-relativistic DM candidates, the later one takes the energy dependence from phase space into account. However, the both treatments are not appropriate for an energetically boosted DM candidate whose kinetic energy is much larger than its mass such that the momentum transfer $q$ cannot be neglected. We consider the relativistic kinematics throughout this work.
As shown in Fig.~\ref{fig:DMflux}, the flux distribution exhibits a significant enhancement at the large $T_\chi$ range with increasing $m_{A^\prime}$.
Note that various recoil flux curves intersect at $T_\chi=(\alpha m_e)^2/(2m_\chi)$, and the recoil flux distribution of the constant $\overline{|\mathcal{M}|^{2}}$ slightly deviates from that of the constant $\sigma_{\chi e}$ when $2 m_\chi T_\mathrm{CR} > (m_e + m_\chi)^2$.

It is worth mentioning that the recoil flux distribution is independent of $m_{A^\prime}$ when the dark photon is very heavy ($m_{A^\prime} \gg \sqrt{2 m_{\chi} T_{\chi}}$) or ultralight ($m_{A^\prime} \ll \alpha m_e$). See the red and blue boundaries of the contour. The recoil flux distributions in the above two limits exhibit distinct dependence on $T_\chi$; for example, the recoil flux of ultralight dark photons drops rapidly with $T_\chi$ while the recoil flux of heavy dark photons mildly decreases with $T_\chi$.
The heavy dark photon represents the so-called $Z^\prime$-portal model while the ultralight dark photon the milli-charged DM model~\cite{Holdom:1985ag}.

Equipped with the boosted DM flux, we now discuss the DM direct detection through the DM interaction with the electron in xenon atoms. For the ionization process of $\chi + A \rightarrow \chi + A^+ +e^-$ with the atom $A$ in the $(n,\,l)$ atomic shell, the velocity-averaged differential cross section with respect to the electron recoil energy $E_{R}$ is given by~\cite{Essig:2011nj,Essig:2015cda}
\begin{equation}
\frac{d \langle\sigma_{ion}^{nl} v\rangle}{d \ln E_{R}}=\frac{\bar{\sigma}_{e}}{8 \mu_{\chi e}^{2}} \int q d q\left|F_{D M}(q)\right|^{2}\left|f_{ion}^{nl}(k^{\prime}, q)\right|^{2} \eta\left(E_{\min }\right),
\label{eq:dcs}
\end{equation}
where $F_{D M}$ is the DM form factor, $\eta$ denotes the mean inverse speed function and $\left|f_{ion}^{nl}(k^{\prime}, q)\right|^{2}$ represents the ionization form factor for an electron with initial state $(n,l)$ and final state with momentum $k^{\prime}=\sqrt{2 m_e E_R}$.
In the case of boosted DM,   the DM form factor $F_{D M}$ is
\begin{eqnarray}
&&|F_{D M}(q)|^{2}
 =\frac{\left(\alpha^2 m_e^2+m_{A^\prime}^{2}\right)^2}{\left(2 m_{e} E_{R}+m_{A^\prime}^{2}\right)^{2}} \nonumber\\
&\times& \frac{2 m_{e}\left(m_{\chi}+T_{\chi}\right)^{2}-E_{R}\left(\left(m_{\chi}+m_{e}\right)^{2}+2 m_{e} T_{\chi}\right)+m_{e} E_{R}^{2}}{2 m_{e} m_{\chi}^{2}}.\nonumber\\
\label{eq:fdm}
\end{eqnarray}
In the non-relativistic limit, $T_\chi,E_{R}\ll m_e$, it reproduces the form factor without CR-boost effects, i.e.,
\begin{equation}
|F_{D M}(q)|^{2}=\left(\frac{\alpha^2 m_e^2 + m_{A^\prime}^{2}}{q^2 + m_{A^\prime}^{2}}\right)^2.
\end{equation}
The mean inverse speed function $\eta$ is replaced by~\cite{An:2017ojc}
\begin{equation}
\eta\left(E_{\min}\right)=\int_{E_{\min }} dE_\chi \Phi_{\rm halo}^{-1} \frac{m_\chi^2}{p E_\chi}\frac{d\Phi_\chi}{dT_\chi}\,,
\label{eq:eta}
\end{equation}
where $\Phi_{\rm halo}= n_\chi \bar{v}_\chi$ is the background DM flux in Galactic halo with $\bar{v}_\chi$ being the corresponding average velocity. Here, $E_{\min}$ is the minimal DM energy to trigger electron with recoil energy $E_R$. Similarly, Eq.~(\ref{eq:eta}) reproduces the conventional expression
\begin{equation}
\eta(v_{\min })=\int_{v_{\min }} \frac{1}{v} f(v) d^3 v
\end{equation}
in the non-relativistic limit.
The ionization form factor $\left|f_{ion}^{nl}(k^{\prime}, q)\right|^{2}$ is calculated by using the Roothaan-Hartree-Fock radial wavefunction~\cite{Bunge:1993jsz} for initial electron state and applying plane wave approximation for final state. For initial electron state, we take into account contributions from $(5p^6,5s^2,4d^{10},4p^6,4s^2)$ xenon electron shells. The differential ionization rate is obtained by multiplying Eq.~(\ref{eq:dcs}) with background DM flux $\Phi_{\rm halo}$, the number of target atoms $N_T$, and sum over different electron shells,
\begin{equation}
\frac{d R_{ion}}{d \ln E_R}= N_T\Phi_{\rm halo} \sum_{nl} \frac{d \langle\sigma_{ion}^{nl} v\rangle}{d \ln E_R}.
\end{equation}

\begin{figure}
\centering
\includegraphics[width=0.45\textwidth]{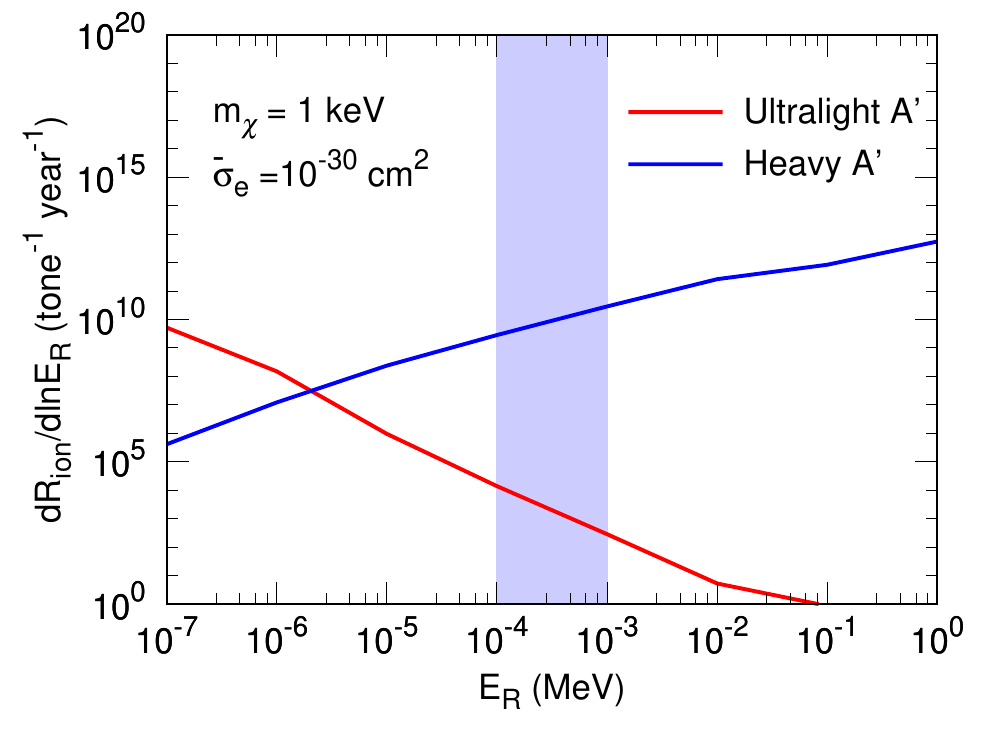}
\caption{Recoil spectra  of electrons for benchmark DM mass $m_\chi=1$ keV with scattering cross section $\bar{\sigma}_e=10^{-30}~{\rm cm}^2$. Here we simply consider recoil electron from $5s$ state for demonstration.}
\label{fig:dRdlnE}
\end{figure}

Figure~\ref{fig:dRdlnE} shows the ionization rate as a function of the electron recoil energy $E_R$ (in unit of ${\rm tonne}^{-1}~{\rm year}^{-1}$) for both ultralight (red) and heavy (blue) dark photons with the choices of $m_\chi=1~{\rm keV}$ and  $\bar{\sigma}_e=10^{-30}~\mathrm{cm}^2$.
The vertical band represents the order of magnitude of energy coverage for current xenon experiments.
The ultralight dark photon prefers to produce electrons with small recoil energy; however, the heavy dark photon is likely to generate electrons with large recoil energy. The distinct difference follows from the energy dependence in the distribution of $d \sigma_{\chi e}/d T_{\chi}$ and the DM form factor $F_{D M}(q)$. It implies that one might distinguish between the dark photon and the heavy dark photon from the recoil energy spectrum of the ionized electron when the background is well understood.

The recoiling electron are then converted into scintillation ($S1$) and ionization ($S2$) signal in liquid xenon experiments, and the observable is the number of photonelectrons (PE). We consider $S2$ signal hereafter as  the XENON100 and XENON1T collaborations release the data sets that are based only on the ionization signal~\cite{Aprile:2016wwo,Aprile:2019xxb}. The event spectrum can be schematically written as following:
\begin{equation}
\frac{dN}{dS2}=T_{\rm exp}\cdot \varepsilon_{S2} \sum_{nl} \int dE_R\,{\rm pdf}\left(S2|\Delta E_e\right)\frac{dR^{nl}_{ion}}{dln E_R}\,,
\label{eq:signal}
\end{equation}
where $T_{\rm exp}$ is the exposure of detector and $\varepsilon_{S2}$ is the efficiency of triggering and accepting the $S2$ signal. For a given deposit energy $\Delta E_e= E_R+|E^{nl}_B|$ with $|E^{nl}_B|$ the binding energy of $(n,l)$ shell, the conversion probability of $S2$ is ${\rm pdf}\left(S2|\Delta E_e\right)$, which is modeled as follows~\cite{Essig:2012yx,Essig:2017kqs}.  The number of primary quanta produced at the interaction point is $n^{(1)}_Q = {\rm Floor}(E_R/W)$ with $W=13.8$ eV, and $n^{(1)}_Q$  is divided into $n_e$ observable ionized electrons escaping from interaction point and $n_\gamma$ unobservable scintillation photons. The fiducial value of the fraction of primary quanta identified as electrons is chosen as $f_e\simeq 0.83$. In addition, in the case of the DM candidate ionizes an inner shell electron, the secondary quanta is  produced by subsequent electron transitions from outer to inner shell. The number of the secondary quanta is $n^{(2)}_Q={\rm Floor}((E_i-E_j)/W)$ where $E_i$ denotes the binding energy of the $i$th shell.  The production number of secondary electrons follows a binomial distribution with $n_Q^{(1)}+n_Q^{(2)}$ trials and the success probability $f_e$. Finally, the number of PE converted from electrons (with total number $n_e=n^{(1)}_e+n^{(2)}_e$) is described by a gaussian distribution with mean value $n_e \mu$ and width $\sqrt{n_e} \sigma$. The parameters are chosen as $\mu=19.7~(11.4)$ and $\sigma=6.9~(2.8)$~\cite{Aprile:2016wwo,Aprile:2019dme}.

\begin{figure}[b!]
\centering
\subfigbottomskip=-100pt
\subfigcapskip=-10pt
\includegraphics[scale=0.55]{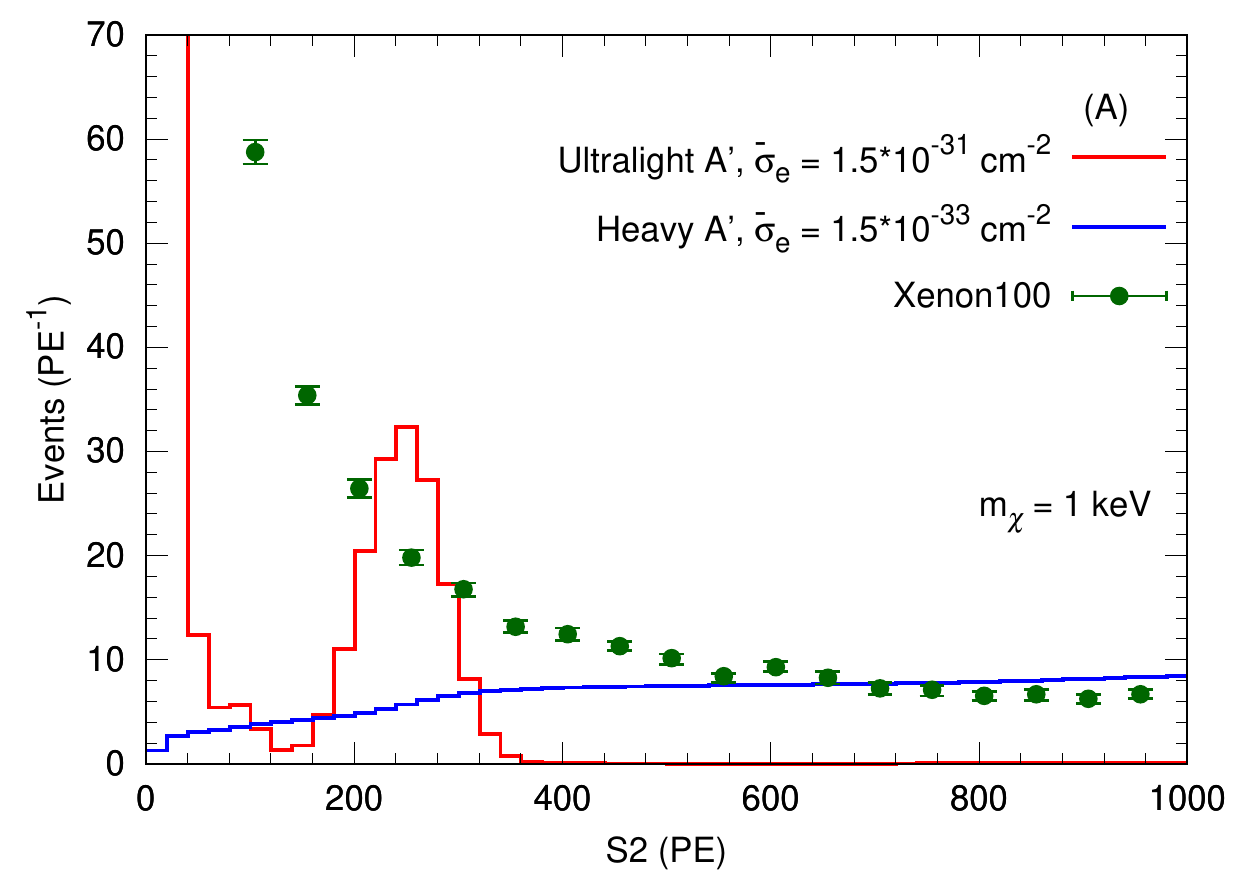}
\includegraphics[scale=0.55]{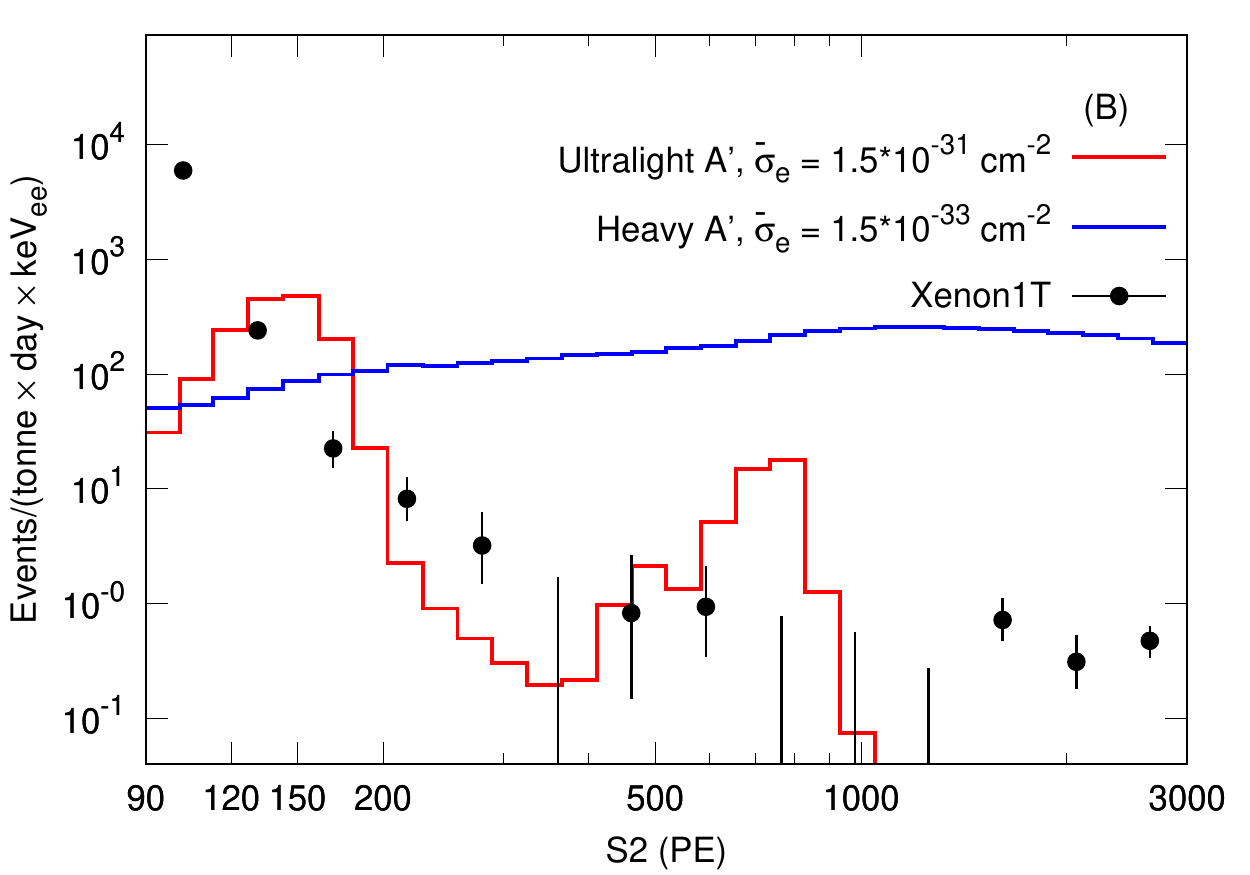}
\caption{Example of expected PE spectra for DM-electron scattering in XENON100 \textbf{(A)} and XENON1T \textbf{(B)} experiments, for both ultralight and heavy mediator cases. Signal spectra are shown for $m_\chi=1$ keV with scattering cross section $\bar{\sigma}_e=1.5\times 10^{-31}~{\rm cm}^2$ ($1.5\times10^{-33}~{\rm cm}^2$) in ultralight (heavy) mediator case.}
\label{fig:pesp}
\end{figure}

\begin{figure*}[tbp!]
\centering
\subfigbottomskip=-100pt
\subfigcapskip=-10pt
\subfigure{ \includegraphics[width=0.47\textwidth]{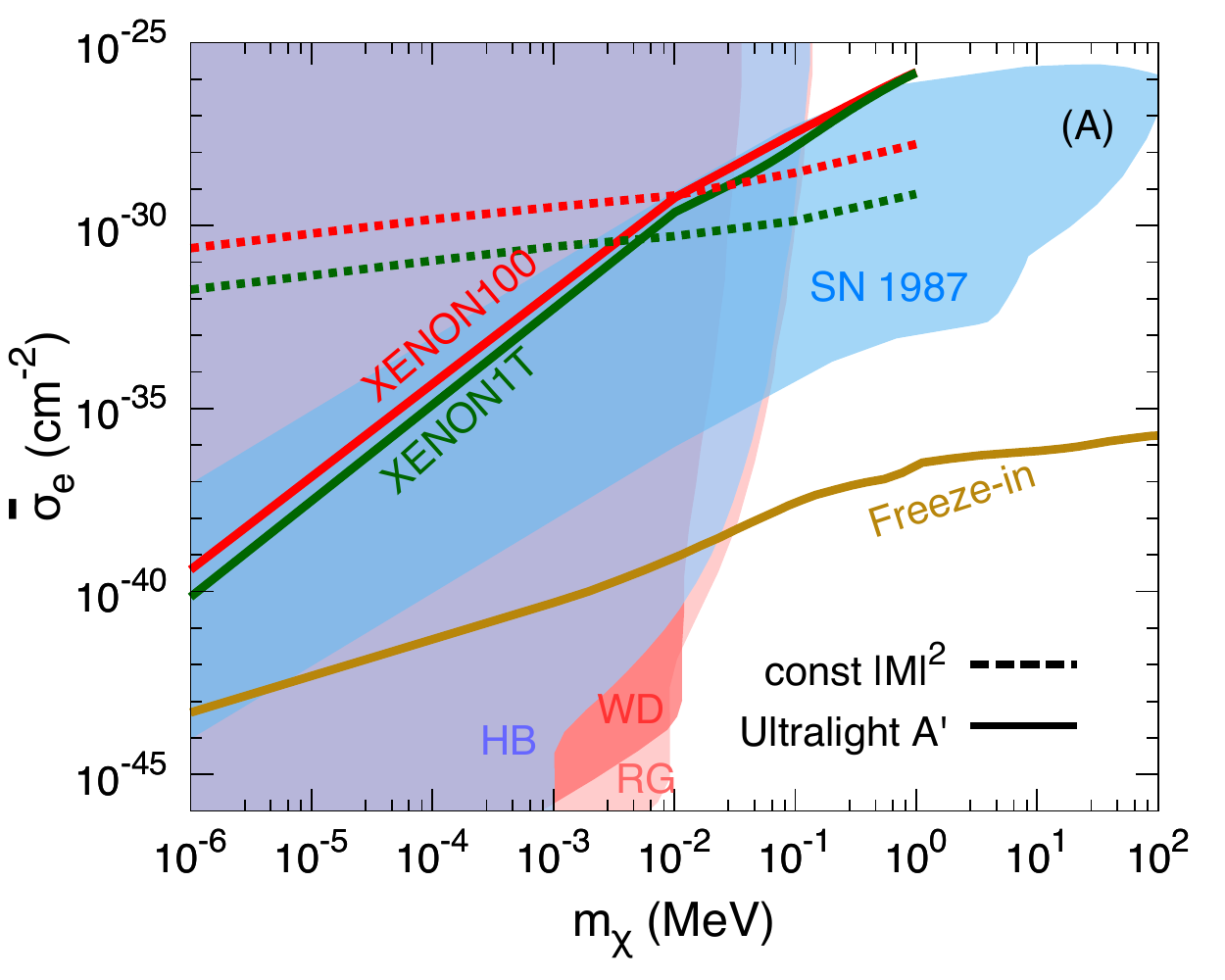}}
\subfigure{ \includegraphics[width=0.47\textwidth]{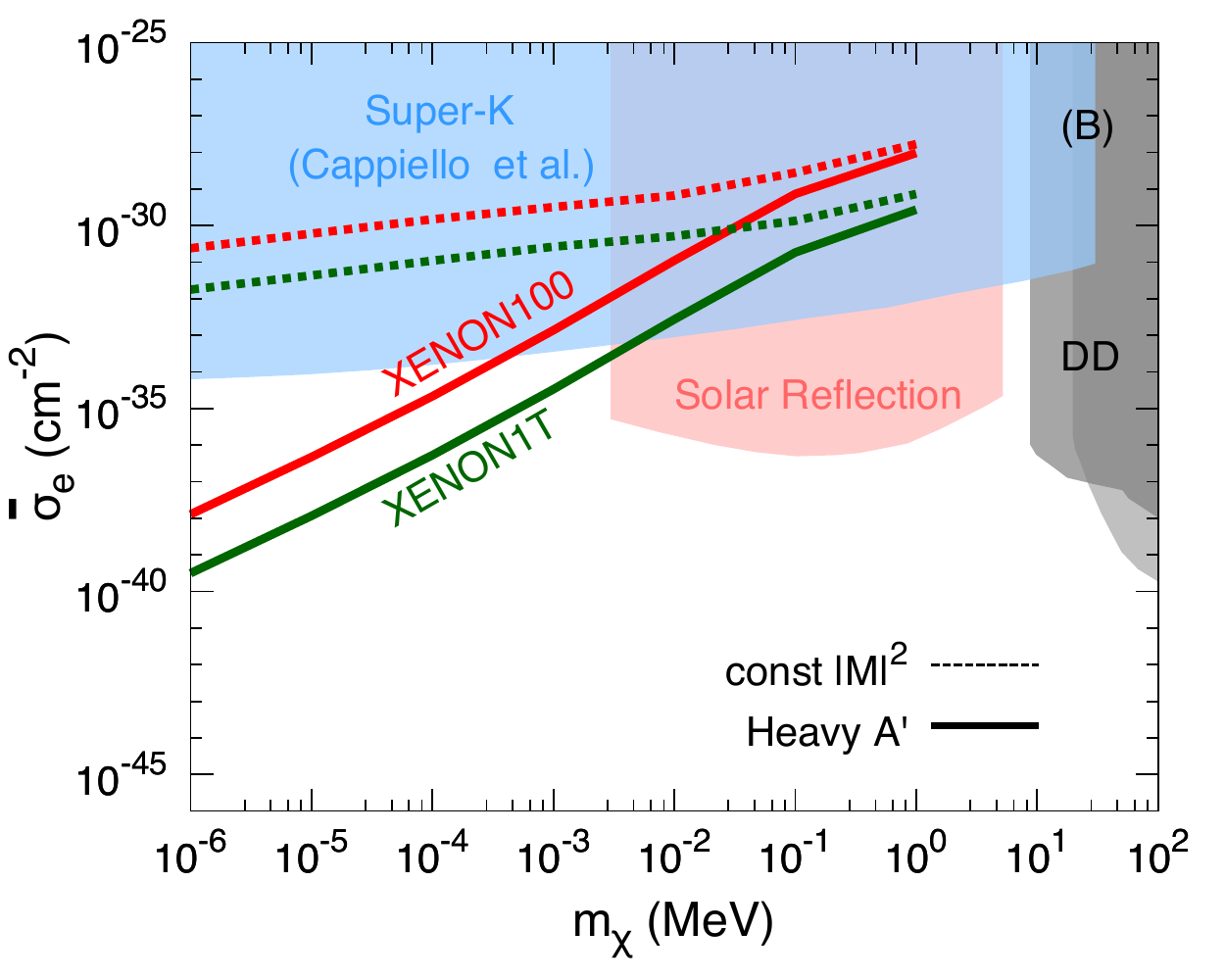}}
\caption{ \textbf{(A)}: exclusion limits in the $m_\chi$-$\bar{\sigma}_e$ plane from the XENON100 data (red-solid) and the XENON1T data (green-solid) for ultralight mediator scenario. For comparison, corresponding limits for constant $\overline{|\mathcal{M}|^{2}}$ are presented by dashed lines. Also shown are cooling constraints from supernovae 1987A (``SN 1987'')~\cite{Chang:2018rso}, energy loss of Red-Giant and Horizontal-Branch stars (``RG \& HB"), as well as white dwarfs (``WD'')~\cite{Vogel:2013raa}; we also plot parameter region where DM obtains the correct relic abundance via freeze-in mechanism~\cite{Essig:2011nj,Dvorkin:2019zdi}. \textbf{(B)}: exclusion limits for heavy mediator scenario. We also plot constraints from Super-Kamiokande neutrino experiment (``Super-K'')~\cite{Cappiello:2019qsw}, solar reflection (``Solar Reflection'')~\cite{An:2017ojc}, as well as previous limits from the XENON10 and the XENON100 direct detections (``DD'')~\cite{Essig:2012yx,Essig:2017kqs}.}   \label{fig:limits}
\end{figure*}

We derive the limits of $\bar{\sigma}_e$ imposed by the XENON100 data~\cite{Aprile:2016wwo} ($T_{\rm exp}=30~{\rm kg-years}$) and by the XENON1T data~\cite{Aprile:2019xxb} (effective $T_{\rm exp}=22~{\rm tonne-days}$), using the same bin steps. We choose the detection efficiency as $\varepsilon_{S2}=1$ for simplicity and obtain the limits by demanding that signal does not exceed $1\sigma$ upper bound in each bin. Figure~\ref{fig:pesp} presents benchmark signal spectra versus PE for the ultralight and heavy mediator cases.

Figure \ref{fig:limits}(A) shows the exclusion limits in the $m_\chi$-$\bar{\sigma}_e$ plane for the case of a ultralight mediator, derived from the XENON100 data (red) and the XENON1T data (green). The acceleration mechanism greatly enhances the discovery potential of direct detection experiments on a light DM candidate. For comparison we also plot the parameter region for the freeze-in DM (brown curve)~\cite{Essig:2011nj,Dvorkin:2019zdi}. Even though the parameter space of freeze-in DM is well below the current direct detection sensitivity,  it can be reached when large experimental exposures are achieved. For example,
the experimental exposure of 30 tonne-years can probe the signal region of freeze-in DM with $m_\chi\sim 1$ eV when the background is fully controlled.
In addition, the DM with a ultralight mediator (or equivalent milli-charged DM) can also be constrained by astrophysical observations from supernova cooling and stellar energy loss~\cite{Chang:2018rso,Vogel:2013raa}. The bounds from the direct detection experiments are comparable to those astrophysical constraints.

Figure~\ref{fig:limits}(B) displays the exclusion limit of $\bar{\sigma}_e$ for the case of a heavy mediator.  We also plot the limits from Super-Kamiokande neutrino experiment~\cite{Cappiello:2019qsw}, solar reflection~\cite{An:2017ojc}, and the direct detection without CR-DM scattering effect~\cite{Essig:2017kqs}. After considering the CR-DM effect, the direct detection experiments have a better sensitivity in the sub-keV mass region.

In summary, we studied the effect of boosted DM on DM-electron direct detections and demonstrate that the current data from liquid noble gas experiments is sensitive to light DM candidates in the range of sub-MeV. More importantly, the energy dependence in cross section plays a crucial role in improving the exclusion limits, e.g., the recoil spectra increase with recoil energy for heavy mediator case while decrease with recoil energy for ultralight mediator. Such opposite energy dependences imply that the neutrino experiments such as Super-K are more powerful for heavy mediator due to their much larger acceptance volume and higher energy coverage~\cite{Bays:2011si}. On the other hand, direct detection has more advantage on ultralight mediator. Such two kind of experiments are complementary to each other.

The CR boosted DM mechanism has very rich phenomenologies. For example, it is interesting to investigate boosted DM flux coming from the Galactic center which possesses high DM density and CR flux. One also expects that the morphology of signal resulted from the Galactic center is different from that originated from local interstellar~\cite{Carlson:2015daa}. Moreover, light DM with significant CR acceleration and heavy DM ($m_\chi \gtrsim 10~\mathrm{MeV}$) with negligible CR acceleration could potentially produce degenerate signal; therefore, discrimination of such two kinds of scenarios in both model independent and model specific way is an intriguing issue~\cite{cao:2019cr}. The boosted mechanism might explain or be constrained by the recoiled energy spectrum of electrons recently reported by the XENON1T collaboration~\cite{Aprile:2020tmw}.   

\noindent {\bf Acknowledgments.~}
We thank Tien-Tien Yu and Su-jie Lin for helpful discussions. The work is supported in part by the National Science Foundation of China under Grant Nos. 11725520, 11675002 and 11635001. QFX is also supported by the China Postdoctoral Science Foundation under Grant No. 8206300015.

\bibliography{Ref}

\newpage
\onecolumngrid
\input{supplement.tex}

\end{document}

%% file: supplement.tex

\newpage
\begin{center}
{\bf SUPPLEMENTAL MATERIALS}
\end{center}

The supplemental materials provide additional details to various results presented in the main text. Some of the results can be applied to other light DM models.

\section{Calculation of CR electron flux}

\begin{figure}[ht!]
\centering
\includegraphics[width=0.46\textwidth]{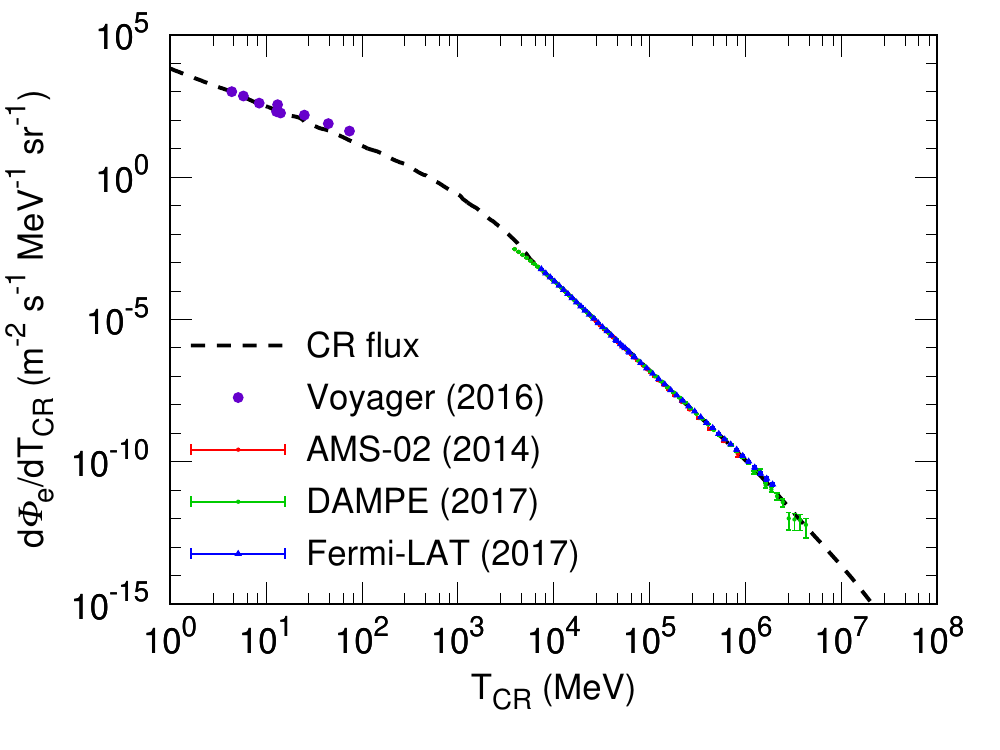}
\caption{Local interstellar flux of CR electrons as a function of electron kinetic energy $T_{\rm CR}$ with the data sets from Voyager 1~\cite{Cummings:2016pdr}, AMS-02~\cite{Aguilar:2014mma} and DAMPE~\cite{Ambrosi:2017wek} measurements. For completeness, we also present Fermi-LAT~\cite{Abdollahi:2017nat} measurement. }
\label{fig:CRflux}
\end{figure}

In order to obtain accurate DM recoil flux, the reliable inputs of electron CR flux are in order. The observed CR electron spectrum at the Earth extend many orders of magnitude energy, ranging from GeV to TeV. Such energetic CR electrons are easy to accelerate a fraction of DM particles to relativistic speeds. The flux of CR electrons is obtained by solving the diffusion equation with a widely used galactic CR propagation model. The flux is also modulated periodically according to the solar activity due to interactions of CR electrons with the heliosphere magnetic field. As a result, the CR spectrum observed at the Earth is different from the one in the interstellar. Such solar modulation is more important for low energy CR electrons and is negligible for energy above several GeV. The unmodulated local interstellar spectra of CR electrons has been measured by Voyager 1 collaboration which covers energy range with $2.7-74$ MeV~\cite{Cummings:2016pdr}.
For high energy CR electrons, AMS-02~\cite{Aguilar:2014mma} and DAMPE~\cite{Ambrosi:2017wek} measurements cover energy ranges from $1$ GeV to $4.6$ TeV. We use the {\tt GALPROPv54}~\cite{Strong:1998pw,Moskalenko:1997gh}  to obtain the best-fit flux for AMS-02 and DAMPE data sets, and combine the best estimation of Voyager 1 data~\cite{Potgieter:2015jxa}. Corresponding local interstellar spectrum of CR electrons  is shown in Fig.~\ref{fig:CRflux} with measurements.

\section{Derivation of the CR-DM differential scattering cross section}

In the CR-DM scattering, the initial DM particles are treated as being at rest since
their typical velocities ($v\sim 10^{-3}$) are negligible compare to the velocities of incoming CR electrons. The recoil energy of DM for a given CR kinetic energy $T_{\rm CR}$ can be calculated from standard relativistic kinematics of 2-body scattering process~\cite{goldstein:2001} and are given as
\begin{equation}
T_\chi=T_\chi^{\max}\frac{(1-\cos \theta_{\rm CM})}{2},\quad T_\chi^{\rm max}=\frac{2 m_\chi T_{\rm CR}(T_{\rm CR}+2m_e)}{(m_e+m_\chi)^2+2T_{\rm CR}m_\chi}\,,
\label{eq:A1}
\end{equation}
where $\theta_{\rm CM}$ is the center-of-mass scattering angle. From above equation, $\theta_{\rm CM}$ and $T_\chi$ are related as
\begin{equation}
\frac{d\cos\theta_{\rm CM}}{d T_\chi}=-\frac{2}{T_\chi^{\max}}\,,
\label{eq:A2}
\end{equation}
which allows us to translate the variable in differential cross section from solid angle $d \Omega$ to DM kinetic energy $d T_\chi$ via
\begin{equation}
\frac{d \sigma_{\chi e}}{d T_\chi} =\frac{d \sigma_{\chi e}}{d \Omega}\cdot\frac{d \Omega}{d T_\chi} =\frac{\overline{|\mathcal{M}|^{2}}}{16 \pi s} \frac{1}{T_\chi^{\max}}\,,
\label{eq:A3}
\end{equation}
where $\overline{|\mathcal{M}|^{2}}=\frac{1}{4} \sum_{\rm spins}|\mathcal{M}|^{2}$ is the squared DM-electron scattering matrix element, averaged over initial and summed over final spin states. Using Eq.~(\ref{eq:A3}) and expressions of Mandelstam variables
\begin{equation}
\left\{\begin{array}{l}{s=(m_\chi + m_e)^2+2 m_\chi T_{\rm CR}}\,, \\ {t=-2 m_{\chi} T_\chi = -q^2}\,, \\ {u=(m_{\chi}-m_e)^2-2m_{\chi}(T_{\rm CR}-T_\chi)}\,, \end{array}\right.
\label{eq:A4}
\end{equation}
one can drive formula of $d\sigma_{\chi e}/d T_{\chi}$ for a given interaction. As below, we list expressions of $d\sigma_{\chi e}/d T_{\chi}$ for some typical interactions, which are widely used in light DM model:
\begin{itemize}
\item {Scalar interaction: $\mathcal{L} \supset g_\chi \overline{\chi} \chi \phi + g_{\rm SM}\overline{f} f \phi$\,,
\begin{eqnarray}
\overline{|\mathcal{M}|^2} &=& g_{\chi}^{2} g_{\rm SM}^{2} \frac{4 m_{\chi}\left(2 m_{\chi}+T_\chi\right)\left(2 m_e^2 + m_{\chi}T_\chi\right)}{\left(2 m_{\chi} T_{\chi}+m_\phi^{2}\right)^{2}}\,, \\
\frac{d \sigma_{\chi e}}{d T_{\chi}} &=& g_\chi^{2} g_{\rm SM}^{2} \frac{\left(2 m_{\chi} + T_{\chi}\right)\left(2 m_e^2 + m_{\chi}T_\chi\right)}{8 \pi\left(2 m_e T_{\rm CR} + T_{\rm CR}^2\right)\left(2 m_{\chi} T_{\chi}+m_\phi^{2}\right)^{2}}\,.
\end{eqnarray} }
\item {Vector interaction: $\mathcal{L} \supset g_\chi \overline{\chi} \gamma^\mu \chi A^\prime_\mu + g_{\rm SM}\overline{f} \gamma^{\mu} f  A^\prime_\mu$\,,
\begin{eqnarray}
\overline{|\mathcal{M}|^2} &=& g_{\chi}^{2} g_{\rm SM}^{2} \frac{8 m_{\chi}\left(2 m_{\chi}\left(m_e+T_{\rm CR}\right)^{2}-T_{\chi}\left(\left(m_e+m_{\chi}\right)^{2}+2 m_{\chi} T_{\rm CR}\right)+m_{\chi} T_{\chi}^{2}\right)}{\left(2 m_{\chi} T_{\chi}+m_{A^\prime}^{2}\right)^{2}}\,, \label{eq:A5}\\
\frac{d \sigma_{\chi e}}{d T_{\chi}} &=& g_\chi^{2} g_{\rm SM}^{2} \frac{2 m_{\chi}\left(m_e+T_{\rm CR}\right)^{2}-T_{\chi}\left(\left(m_e+m_{\chi}\right)^{2}+2 m_{\chi} T_{\rm CR}\right)+m_{\chi} T_{\chi}^{2}}{4 \pi\left(2 m_e T_{\rm CR}+T_{\rm CR}^{2}\right)\left(2 m_{\chi} T_{\chi}+m_{A^\prime}^{2}\right)^{2}}\,.
\label{eq:A6}
\end{eqnarray}}
\item {Axial-vector interaction: $\mathcal{L} \supset g_\chi \overline{\chi} \gamma^\mu \gamma^5 \chi A^\prime_\mu + g_{\rm SM}\overline{f} \gamma^{\mu} \gamma^5 f  A^\prime_\mu$\,,
\begin{eqnarray}
\overline{|\mathcal{M}|^2} &=& g_{\chi}^{2} g_{\rm SM}^{2} \frac{8 m_{\chi}\left(2 m_{\chi}\left(\left(m_e+T_{\rm CR}\right)^{2}+2 m_e^{2}\right)+T_{\chi}\left(\left(m_e-m_{\chi}\right)^{2}-2 m_{\chi} T_{\rm CR}\right)+m_{\chi} T_{\chi}^{2}\right)}{\left(2 m_{\chi} T_{\chi}+m_{A^\prime}^{2}\right)^{2}}\,, \\
\frac{d \sigma_{\chi e}}{d T_\chi} &=& g_\chi^{2} g_{\rm SM}^{2} \frac{2 m_{\chi}\left(\left(m_e+T_{\rm CR}\right)^{2}+2 m_e^{2}\right)+T_{\chi}\left(\left(m_e-m_{\chi}\right)^{2}-2 m_{\chi} T_{\rm CR}\right)+m_{\chi} T_{\chi}^{2}}{4 \pi\left(2 m_e T_{\rm CR}+T_{\rm CR}^{2}\right)\left(2 m_{\chi} T_{\chi}+m_{A^\prime}^{2}\right)^{2}}\,.
\end{eqnarray}}
\end{itemize}

For the purpose of this paper, we concentrate on vector interaction, while the limits for other interactions can be obtained in a straightforward way by using our calculation procedures. The DM-electron elastic scattering cross section is conventionally normalized to $\bar{\sigma}_{e}$ with following definitions~\cite{Essig:2011nj}:
\begin{eqnarray}
\overline{|\mathcal{M}_{\rm free}|^2} &=& \overline{|\mathcal{M}_{\rm free}\left(\alpha m_{e}\right)|^2}\times\left|F_{D M}(q)\right|^2 \,, \label{eq:A7}\\
\bar{\sigma}_{e} &=& \frac{\mu_{\chi e}^2\overline{|\mathcal{M}_{\rm free}\left(\alpha m_{e}\right)|^2}}{16 \pi m_{\chi}^{2} m_{e}^2}\,,
\label{eq:A8}
\end{eqnarray}
where $\mu_{\chi e}$ is the DM-electron reduced mass, $\mathcal{M}_{\rm free}\left(\alpha m_{e}\right)$ is corresponding matrix element for momentum transfer at reference value $q=|\boldsymbol{q}|=\alpha m_e$. The DM form factor, $F_{D M}(q)$, encapsulates all remaining energy dependence of the interaction.
With the notation of Eq.~(\ref{eq:A8}), the DM-electron reference cross section for benchmark model in Eq.~(\ref{eq:lag}) is given by
\begin{eqnarray}
\overline{|\mathcal{M}_{\rm free}\left(\alpha m_{e}\right)|^2} &=& \frac{16 g_\chi^{2} g_{\rm SM}^{2} m_{e}^{2} m_{\chi}^{2}}{\left(\alpha^2 m_e^2+m_{A^\prime}^{2}\right)^{2}}\,, \label{eq:A9}\\
\bar{\sigma}_{e} &=& \frac{g_\chi^{2} g_{\rm SM}^{2} \mu_{\chi e}^{2}}{\pi \left(\alpha^2 m_e^2+m_{A^\prime}^{2}\right)^{2}}\,.
\label{eq:A10}
\end{eqnarray}
Combining Eqs.~(\ref{eq:A6}) and (\ref{eq:A10}) then gives expression of $d\sigma_{\chi e}/d T_{\chi}$ in Eq.~(\ref{eq:dsdT})
\begin{eqnarray}
\frac{d \sigma_{\chi e}}{d T_{\chi}}
&=& \bar{\sigma}_e \frac{\left(\alpha^2 m_e^2+m_{A^\prime}^{2}\right)^{2}}{\mu_{\chi e}^{2}} \frac{2 m_{\chi}\left(m_e+T_{\rm CR}\right)^{2}-T_{\chi}\left(\left(m_e+m_{\chi}\right)^{2}+2 m_{\chi} T_{\rm CR}\right)+m_{\chi} T_{\chi}^{2}}{4 \left(2 m_e T_{\rm CR}+T_{\rm CR}^{2}\right)\left(2 m_{\chi} T_{\chi}+m_{A^\prime}^{2}\right)^{2}} \nonumber\\
&\simeq& \bar{\sigma}_e
\begin{cases} \frac{2 m_{\chi}\left(m_e+T_{\rm CR}\right)^{2}-T_{\chi}\left(\left(m_e+m_{\chi}\right)^{2}+2 m_{\chi} T_{\rm CR}\right)+m_{\chi} T_{\chi}^{2}}{4 \mu_{\chi e}^{2} \left(2 m_e T_{\rm CR}+T_{\rm CR}^{2}\right)},\quad \ \ &
  {\rm heavy}~A^\prime\\
 \frac{\alpha^4 m_e^4}{16 m^2_{\chi} T^2_{\chi}}\frac{2 m_{\chi}\left(m_e+T_{\rm CR}\right)^{2}-T_{\chi}\left(\left(m_e+m_{\chi}\right)^{2}+2 m_{\chi} T_{\rm CR}\right)+m_{\chi} T_{\chi}^{2}}{\mu_{\chi e}^{2} \left(2 m_e T_{\rm CR}+T_{\rm CR}^{2}\right)}, \quad \ \ & {\rm ultralight}~A^\prime
\end{cases}\,.
\label{eq:A11}
\end{eqnarray}
Finally, from Eqs.~(\ref{eq:A3}) and (\ref{eq:A8}), one can easily drive $d\sigma_{\chi e}/d T_{\chi}$ corresponding to constant scattering cross section ($\overline{|\mathcal{M}|^{2}}/(16 \pi s)\equiv \bar{\sigma}_{e}$) and constant matrix element. Which are respectively given as
\begin{equation}
\frac{d\sigma_{\chi e}}{dT_\chi}=\bar{\sigma}_{e}
\begin{cases}\frac{1}{T^{\max}_\chi} ,\quad \ \ &
  {\rm constant}\,\sigma_{\chi e}\\
\frac{\left(m_{\chi}+m_{e}\right)^{2}}{\left(m_{\chi}+m_{e}\right)^{2}+2 m_{\chi} T_{\rm CR}}\frac{1}{T^{\max}_\chi}, \quad \ \ & {\rm constant}\, \overline{|\mathcal{M}|^{2}}
\end{cases}\,.
\label{eq:A12}
\end{equation}

\begin{figure}[ht!]
\centering
\subfigbottomskip=-100pt
\subfigcapskip=-15pt
\subfigure{\includegraphics[width=0.4\textwidth]{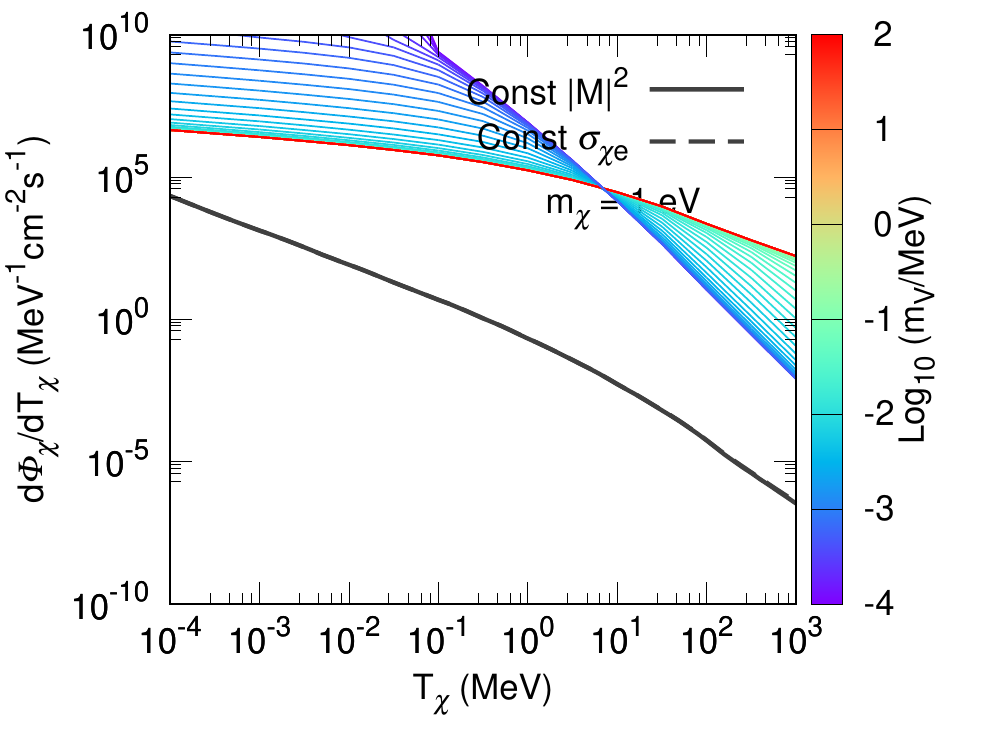}}
\subfigure{\includegraphics[width=0.4\textwidth]{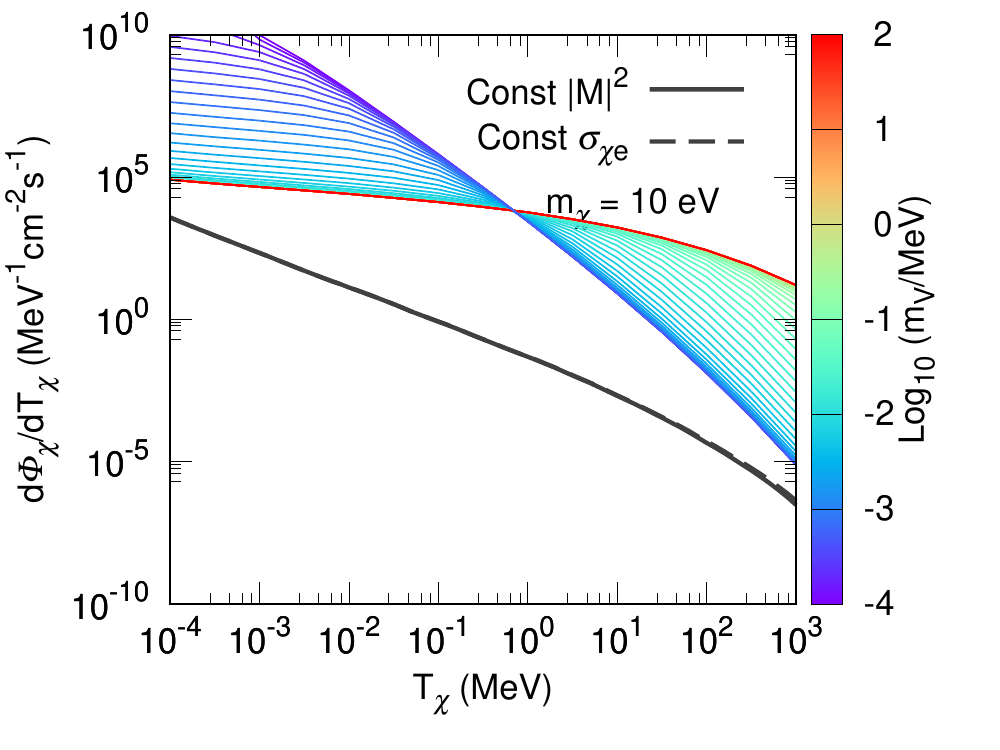}}
\subfigure{\includegraphics[width=0.4\textwidth]{figs/DM_flux.pdf}}
\subfigure{\includegraphics[width=0.4\textwidth]{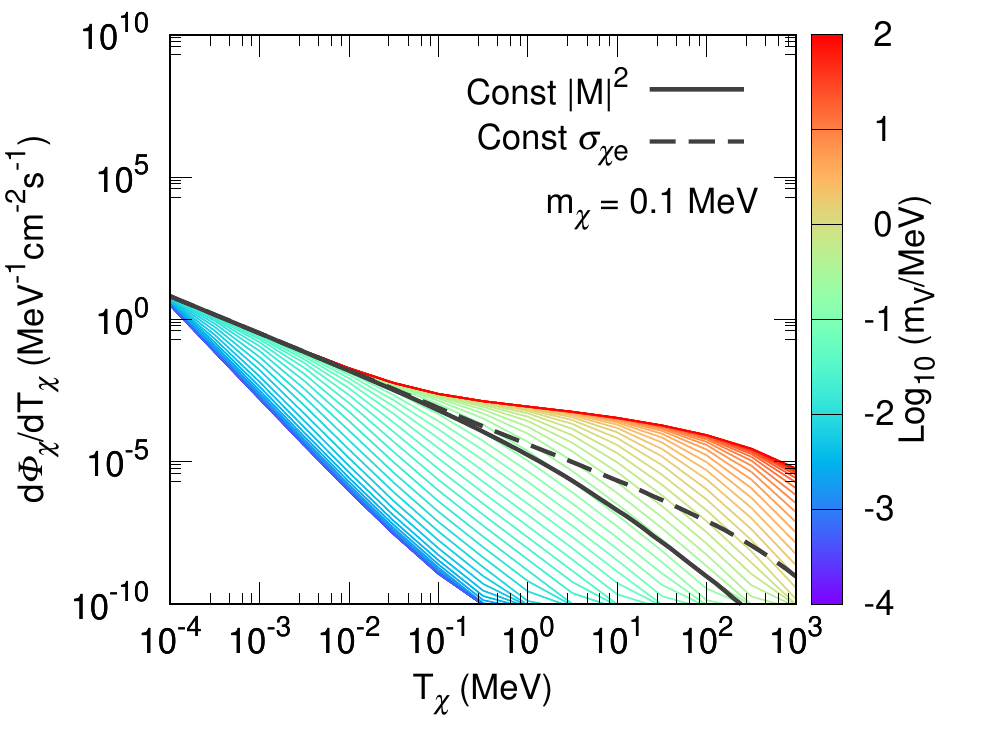}}
\caption{DM recoil fluxes for benchmark DM masses $m_\chi=1$ eV,10 eV, 1 keV and $0.1$ MeV with varying mediator mass $m_{A^\prime}$.}
\label{fig:DMflux2}
\end{figure}

Given differential cross section in Eqs.~(\ref{eq:A11}) and (\ref{eq:A12}), we can calculate DM recoil flux as a function of DM kinetic energy according to Eq.~(\ref{eq:DMrecoil}). In Fig.~\ref{fig:DMflux2}, in additional to the $m_\chi=1$ keV recoil flux in the main text, we also present DM recoil fluxes for $m_\chi=1$ eV, 10 eV and 0.1 MeV.

\section{Derivation of the DM-electron scattering cross section}
The cross section of DM particle scattering with electron in a bound state can be derived in a standard way using quantum field theory. In the derivation, one conventionally treats the electron is bounded in a static background potential, which means that the recoiling of atoms is neglected. Under such approximation, the cross section for elastic $2\to 2$ scattering process $\chi(p)+e(k) \rightarrow \chi\left(p^{\prime}\right)+e\left(k^{\prime}\right)$ is given by
\begin{eqnarray}
d \sigma &=& \frac{\overline{|\mathcal{M_{\rm free}}|^{2}}}{v_{\chi e}} \frac{1}{2 k_{0} 2 p_{0}}(2 \pi)^{4} \delta^{4}\left(k+p-k^{\prime}-p^{\prime}\right) \frac{d^{3} \boldsymbol{p}^{\prime}}{(2 \pi)^{3} 2 p_{0}^{\prime}} \frac{d^3 \boldsymbol{k}^{\prime}}{(2 \pi)^{3} 2 k_{0}^{\prime}} \nonumber \\
&=& \frac{\overline{|\mathcal{M_{\rm free}}|^{2}}}{v_{\chi e}} \frac{1}{64 \pi^{2} E_{\chi} E_{\chi}^{\prime} E_e E_e^{\prime}} \frac{1}{(2 \pi)^{3}}\delta\left(\Delta E_\chi-\Delta E_{e}\right)\left[(2 \pi)^{3} \delta^{3}\left(\boldsymbol{k}-\boldsymbol{k}^{\prime}+\boldsymbol{q}\right)\right] d^{3} \boldsymbol{q} d^{3} \boldsymbol{k}^{\prime}\,,
\label{eq:B1}
\end{eqnarray}
where $v_{\chi e}$ is the relative velocity of incoming DM and electron, $\boldsymbol{q}=\boldsymbol{p}-\boldsymbol{p}^{\prime}$ is the
momentum transfer from DM to electron. $\Delta E_\chi$ is the amount of energy lost by DM in the scattering. Notice that for the initial state is bounded electron, one just need to take replacement $(2 \pi)^{3} \delta^{3}\left(\boldsymbol{k}-\boldsymbol{k}^\prime+\boldsymbol{q}\right) \rightarrow |f_{i\to \boldsymbol{k}^\prime}(\boldsymbol{q})|^2$ in Eq.~(\ref{eq:B1}). The atomic form factor, $f_{i\to \boldsymbol{k}^\prime}(\boldsymbol{q})=\sqrt{V}\int d^3 \boldsymbol{r} \psi_i(\boldsymbol{r})\psi^*_{\boldsymbol{k}^\prime}(\boldsymbol{r})e^{i\boldsymbol{q}\cdot\boldsymbol{r}}$,  accounts for transition from initial to final electron states, and $V$ is the volume for wavefunction normalization. To understand the consistency of such replacement, notice that for both initial and final states are free electrons, such atomic form factor reduces to $f_{i\to \boldsymbol{k}^\prime}(\boldsymbol{q})= (2 \pi)^3 \delta^{3}\left(\boldsymbol{k}-\boldsymbol{k}^{\prime}+\boldsymbol{q}\right)$. Here we have included the normalization of the wavefuctions in terms of the volume $V$, and used the large volume limit $(2 \pi)^3 \delta^{3}(0)/V\rightarrow 1$. Then for the ionization process $\chi + A \rightarrow \chi + A^+ +e^-$ in the $(n,\,l)$ atomic shell, Eq.~(\ref{eq:B1}) is recast as
\begin{equation}
d \sigma=\frac{\overline{|\mathcal{M_{\rm free}}|^{2}}}{v_{\chi e}} \frac{1}{64 \pi^{2} E_{\chi} E_{\chi}^{\prime} E_e E_e^{\prime}} \frac{1}{(2 \pi)^{3}} \delta\left(\Delta E_\chi-\Delta E_{e}\right)\left|f_{n l}(\boldsymbol{q})\right|^{2} d^{3} \boldsymbol{q} d^{3} \boldsymbol{k}^{\prime}.
\label{eq:B2}
\end{equation}
Here both initial bounded and recoil electron are non-relativistic, but incoming DM particle could be relativistic in general. The initial bounded electron and recoil electron respectively have energy $E_e =m_e - |E^{nl}_B|$ and $E_e^{\prime}=m_e + E_R$, with $E_R,|E^{nl}_B|\ll m_e$. One can thus take replacement $E_eE_e^{\prime}\simeq m_e^2$ in Eq.~(\ref{eq:B2}). The deposit energy in electron, $\Delta E_{e}$,  is determined by energy conservation $\Delta E_{e} = \Delta E_\chi$  with 
\begin{eqnarray}
\Delta E_\chi &=& E_{\chi}-E_{\chi}^{\prime}=m_{\chi}\left(\sqrt{1+\frac{p^2}{m_{\chi}^{2}}}-\sqrt{1+\frac{p^{2}+q^{2}-2 p q \cos \theta}{m_{\chi}^{2}}}\right)\,, \label{eq:B3}\\
\Delta E_{e} &=& E_e^{\prime}-E_e=|E^{nl}_B|+E_R\,,
\label{eq:B3}
\end{eqnarray}
where $q=|\boldsymbol{q}|$, $p=|\boldsymbol{p}|$. Applying the definitions in Eqs.~(\ref{eq:A7}) and (\ref{eq:A8}), one can simplify Eq.~(\ref{eq:B2}) to
\begin{equation}
d \sigma v_{\chi e} = \frac{\bar{\sigma}_{e}}{4 \pi} \frac{m_{\chi}^{2}}{\mu_{\chi e}^{2}} \frac{\left|F_{D M}(q)\right|^{2}}{E_{\chi} E_{\chi}^{\prime}} \delta\left(\Delta E_\chi-\Delta E_{e}\right)\frac{1}{(2 \pi)^{3}}\left|f_{n l}(\boldsymbol{q})\right|^{2}  d^{3} \boldsymbol{q} d^{3} \boldsymbol{k}^{\prime} .
\label{eq:B4}
\end{equation}
In order to express differential cross section with respect to electron recoil energy $E_{R}$, using the relation $d^{3} \boldsymbol{k}^{\prime}=\frac{1}{2} k^{\prime 3} \,d \ln E_{R}\,d \Omega_{\boldsymbol{\hat{k}}^{\prime}}$, and rewrite $\delta$-function as
\begin{equation}
\delta\left(\Delta E_\chi-\Delta E_{e}\right)=\frac{E_{\chi}^{\prime}}{p q \sin \theta} \delta(\theta) . 
\label{eq:B5}
\end{equation}
Then by taking derivative of $\Delta E_\chi$ in Eq.~(\ref{eq:B3}) with respect to $\theta$, Eq.~(\ref{eq:B4}) is recast to the expected form
\begin{equation}
\frac{d \sigma v_{\chi e}}{d \ln E_{R}} = \frac{\bar{\sigma}_{e}}{8 \mu_{\chi e}^{2}} \int q d q\left|F_{D M}(q)\right|^{2} \left( \frac{2 k^{\prime 3}}{(2 \pi)^{3}}\sum_{\rm{deg}}\left|f_{n l}(\boldsymbol{q})\right|^{2}\right) \left(\frac{m_{\chi}^{2}}{p E_{\chi}}\right)\,.
\label{eq:B6}
\end{equation}
Integrated with the incoming flux of boosted DM $d{\Phi_\chi}/{dT_\chi}$, we finally obtain the velocity averaged differential ionization cross section
\begin{eqnarray}
\frac{d \langle\sigma_{ion}^{nl} v\rangle}{d \ln E_{R}}=\frac{\bar{\sigma}_{e}}{8 \mu_{\chi e}^{2}} \int q d q\left|F_{D M}(q)\right|^{2}\left|f_{ion}^{nl}(k^{\prime}, q)\right|^{2} \eta\left(E_{\min }\right)\,.
\label{eq:B7}
\end{eqnarray}
Here the DM form factor $F_{DM}(q)$ is evaluated by inverting matrix element in Eq.~(\ref{eq:A5})with applying Eqs.~(\ref{eq:A7}) and (\ref{eq:A9}), which reads
\begin{eqnarray}
|F_{D M}(q)|^{2}&=& \frac{\left(\alpha^2 m_e^2+m_{A^\prime}^{2}\right)^2}{\left(2 m_{e} E_R+m_{A^\prime}^{2}\right)^{2}} \frac{2 m_{e}\left(m_{\chi}+T_{\chi}\right)^{2}-E_R\left(\left(m_{\chi}+m_{e}\right)^{2}+2 m_{e} T_{\chi}\right)+m_{e} E_R^{2}}{2 m_{e} m_{\chi}^{2}} \nonumber\\
&\simeq&
\begin{cases} \frac{2 m_{e}\left(m_{\chi}+T_{\chi}\right)^{2}-E_R\left(\left(m_{\chi}+m_{e}\right)^{2}+2 m_{e} T_{\chi}\right)+m_{e} E_R^{2}}{2 m_{e} m_{\chi}^{2}},\quad \ \ &
{\rm heavy}~A^\prime\\
\frac{\alpha^4 m_e^4}{8 m_{e}^2 E_R^2}\frac{2 m_{e}\left(m_{\chi}+T_{\chi}\right)^{2}-E_R\left(\left(m_{\chi}+m_{e}\right)^{2}+2 m_{e} T_{\chi}\right)+m_{e} E_R^{2}}{ m_{e} m_{\chi}^{2}}, \quad \ \ & {\rm ultralight}~A^\prime
\end{cases}\,.
\label{eq:B9}
\end{eqnarray}
It is easy to verify that $|F_{D M}|$ is reduced to conventional expression $|F_{D M}(q)|^{2}=((\alpha^2 m_e^2 + m_{A^\prime}^{2})/(q^2 + m_{A^\prime}^{2}))^2$ in non-relativistic limit, e.g., $T_\chi,E_{R}\ll m_e$.

The generalized $\eta$ function is given by
\begin{equation}
\eta\left(E_{\min}\right)=\int_{E_{\min }} dE_\chi \Phi_{\rm halo}^{-1} \frac{m_\chi^2}{p E_\chi}\frac{d\Phi_\chi}{dE_\chi}\,,
\label{eq:B10}
\end{equation}
where $\Phi_{\rm halo}\equiv n_\chi \bar{v}$ is the background DM flux in the Galactic halo. $E_{\min}$ is the minimum incoming DM energy to produce an electron with recoil energy $E_R$, which is determined by energy conservation $\Delta E_\chi=\Delta E_{e}$ when $\boldsymbol{p}$ and $\boldsymbol{q}$ are parallel $(\cos \theta=1)$ and 
\begin{equation}
p^{\min}=\frac{q}{2\left(1-\Delta E_{e}^{2} / q^{2}\right)}\left(1-\frac{\Delta E_{e}^{2}}{q^{2}} + \frac{\Delta E_{e}}{q} \sqrt{\left(1-\frac{\Delta E_{e}^{2}}{q^{2}}\right)\left(1+\frac{4 m_{\chi}^{2}}{q^{2}}-\frac{\Delta E_{e}^{2}}{q^{2}}\right)}\right)\,.
\label{eq:B11}
\end{equation}
Notice that the flux is related to the velocity distribution $f(\boldsymbol{v})$ with $d\Phi_\chi(\boldsymbol{v})=n_\chi |\boldsymbol{v}| f(\boldsymbol{v})d^3\boldsymbol{v}$.  Eq.~(\ref{eq:B10}) can be expressed as standard form
\begin{eqnarray}
\eta\left(E_{\min }\right) &=& \int_{E_{\min }} \left(\frac{1}{n_\chi \bar{v}}\right)\frac{m_\chi^2}{v E_\chi^2} n_\chi \bar{v} f(v) d^3 v \nonumber\\
&=& \int_{E_{\min }} \frac{m_\chi^2}{v E_\chi^2} f(v) d^3 v\,.
\label{eq:B12}
\end{eqnarray}
Similarly, in the non-relativistic limit, one has
\begin{eqnarray}
p^{\min} &\simeq& \frac{q}{2}\left(1+\frac{\Delta E_e}{q}\frac{2m_{\chi}}{q}\right)=\frac{q}{2}+\frac{m_{\chi}\Delta E_e}{q}\,, \\
v_{\min } &=& \frac{p^{\min}}{m_{\chi}}=\frac{q}{2m_{\chi}}+\frac{\Delta E_e}{q}\,.
\end{eqnarray}
Equation~(\ref{eq:B12}) reduces to standard mean inverse speed function $\eta(v_{\min })=\int_{v_{\min }} \frac{1}{v} f(v) d^3 v$.

Finally, the atomic ionization form factor $\left|f_{ion}^{nl}(k^{\prime}, q)\right|^{2}$ is defined as
\begin{equation}
\left|f_{ion}^{nl}(k^{\prime}, q)\right|^{2}\equiv\frac{2 k^{\prime 3}}{(2 \pi)^{3}} \sum_{\rm deg} \left|f_{n l}(\boldsymbol{q})\right|^{2}\,,
\label{eq:B13}
\end{equation}
where $f_{n l}(\boldsymbol{q})$ is the atomic form factor for $(n,l)$ electron shell. For our interested case, the final electron state is always ionizaed thus
can be taken as a free wavefunction with momentum $k^{\prime}=\sqrt{2 m_e E_R}$. In this case, $f_{n l}(\boldsymbol{q})$ is simplified to
\begin{eqnarray}
\sum_{\rm deg} \left|f_{n l}(\boldsymbol{q})\right|^{2} &=& \sum_{\text {deg}}\left|\langle\boldsymbol{k}^\prime|e^{i\boldsymbol{q}\cdot\boldsymbol{r}}|nlm\rangle \right|^2
= \sum_{\text {deg}}\left|\int d^{3} \boldsymbol{r}e^{-i\boldsymbol{k}\cdot\boldsymbol{r}}\psi_{nlm}(\boldsymbol{r}) \right|^{2} \nonumber \\
&=& \sum_{\rm deg}\left|\chi_{n l}(k) Y_{l m}(\hat{\boldsymbol{k}})\right|^2\,,
\label{eq:B14}
\end{eqnarray}
where we have used the definition of momentum space wavefunction of the initial bounded electron $\psi_{n l m}(\boldsymbol{k})=\int d^{3} \boldsymbol{r} \psi_{n l m}(\boldsymbol{r}) e^{-i \boldsymbol{k} \cdot \boldsymbol{r}} \equiv \chi_{n l}(k) Y_{l m}(\hat{\boldsymbol{k}})$, with the normalization
$\int d^{3} \boldsymbol{k} \left|\psi_{n l m}(\boldsymbol{k})\right|^2 =(2\pi)^3$. $\chi_{n l}(k)$ is the radial wavefunction in momentum space, and $Y_{l m}(\hat{\boldsymbol{k}})$ is the spherical harmonic function which accounts for angular part of the wavefunction. Writing the sum of degenerate states explicitly, we arrive at
\begin{equation}
\sum_{\rm{deg}}\left|f_{n l}(\boldsymbol{q})\right|^{2} = 2 \int d \Omega_{\hat{\boldsymbol{k}}} \sum_{m=-l}^{l}\left|\chi_{n l}(k) Y_{l m}\left(\hat{\boldsymbol{k}}\right)\right|^{2}\,,
\end{equation}
where factor 2 takes account of electron spin. Applying the property of harmonics function
\begin{equation}
\sum_{m=-l}^{l}\left|Y_{l m}\left(\hat{\boldsymbol{k}}\right)\right|^{2} = \frac{2 l+1}{4 \pi}\,,
\end{equation}
and change the integration variable to initial electron momentum $k$ by using $\sin \theta d \theta=k d k/(k^\prime q)$, we obtain the expression of atomic ionization form factor in the literature~\cite{Essig:2011nj}
\begin{eqnarray}
\left|f_{ion}^{nl}(k^{\prime}, q)\right|^{2} &=& \frac{2 k^{\prime 3}}{(2 \pi)^{3}} \left(\frac{2 l+1}{2 \pi} \int d \Omega_{\hat{k}} \left|\chi_{n l}(k)\right|^{2}\right) \nonumber \\
&=& \frac{2 k^{\prime 3}(2 l+1)}{(2 \pi)^{3}} \int \sin \theta d \theta \left|\chi_{n l}\left(\sqrt{k^{\prime 2}+q^{2}-2 k^{\prime} q \cos \theta}\right)\right|^{2} \nonumber \\
&=& \frac{(2 l+1)k^{\prime 2}}{4 \pi^3 q} \int_{|k^\prime-q|}^{|k^\prime+q|} k d k \left|\chi_{n l}(k)\right|^{2}\,.
\end{eqnarray}

\section{Calculation of the radial Roothaan-Hartree-Fock wavefunction}

We here give the detailed computation of the momentum space radial wave function $\chi_{n l}(p)$ for DM-electron elastic scattering, which is used to calculate atomic ionization form factor. $\chi_{n l}(p)$ is obtained by splitting the coordinate space wavefunction $\psi_{n l m}(x)$ into its radial part $R_{nl}(r)$ and its angular part $Y_{l m}(\boldsymbol{\theta},~\boldsymbol{\phi})$, the exact expression is given by~\cite{Kopp:2009et}
\begin{eqnarray}
\chi_{n l}(p) &=& \frac{4 \pi}{2 l+1} \sum_{m} \psi_{n l m}(\mathbf{p}) Y_{l m}\left(\theta_{p}, \phi_{p}\right) \nonumber\\
&=& 4 \pi i^{l} \int d r r^{2} R_{n l}(r) j_{l}(p r).
\label{eq:C1}
\end{eqnarray}
Here, $\boldsymbol{p}$ is a momentum space vector with arbitrary orientation $(\theta_p, \phi_p)$, and $p=|\boldsymbol{p}|$. $P_l(\cos \theta)$ is a Legendre polynomial. To obtain above result, we have used the orthogonality of the spherical harmonics
\begin{equation}
\int_{0}^{\pi} \int_{0}^{2 \pi} Y \operatorname{lm}(\theta, \phi) Y l^{\prime} m^{\prime}(\theta, \phi) \sin \theta d \theta d \varphi=\delta_{l l^{\prime}} \delta_{m m^{\prime}}\,,
\label{eq:C2}
\end{equation}
and the Gegenbauer formula
\begin{equation}
j_{l}(pr)=\frac{(-i)^{l}}{2} \int_{0}^{\pi}-d(\cos \theta) P_{l}(\cos \theta) e^{i pr \cos \theta}\,,
\label{eq:C3}
\end{equation}
which expresses the spherical Bessel function $j_{l}(x)$ with Fourier type integration over Legendre polynomial. In the RHF method, the radial wavefunctions $R_{nl}(r)$ is approximated by a linear combination of Slater-type orbitals~\cite{Bunge:1993jsz}:
\begin{equation}
R_{n l}(r)=\sum_{k} C_{n l k} \frac{\left(2 Z_{l k}\right)^{n_{l k}+1 / 2}}{a_{0}^{3 / 2} \sqrt{\left(2 n_{l k}\right) !}}\left(r / a_{0}\right)^{n_{l k}-1} \exp \left(-\frac{Z_{l k} r}{a_{0}}\right)\,,
\label{eq:C4}
\end{equation}
where $a_{0}$ is the Bohr radius, and the values of coefficients $C_{n l k}$, $Z_{l k}$ and $n_{l k}$ are provided in Ref.~\cite{Bunge:1993jsz}.  Then  $\chi_{n l}(p)$ can be expressed as
\begin{equation}
\chi_{n l}(p) = 4\pi i^l \sum_{k} C_{n l k} \frac{\left(2 Z_{l k}\right)^{n_{l k}+1 / 2}}{ \sqrt{\left(2 n_{l k}\right) !}}
a_0^{1-n_{l k}-3/2}\int_0^\infty dr\,r^{n_{l k}+1}\,e^{-Z_{l k}/a_0}\,j_l(pr)\,.
\label{eq:C5}
\end{equation}
Applying the Hankel transform formula~\cite{wang:1989}
\begin{equation}
\int_0^\infty e^{-at}\,J_\nu(bt)\,t^{\mu-1}dt=\frac{\Gamma(\mu+\nu)}{a^{\mu+\nu}\Gamma(\nu+1)}\left(\frac{b}{2}\right)^\nu \,_{2} F_{1}\left[\frac{\mu+\nu}{2}, \frac{\mu+\nu+1}{2}, \nu+1, -\frac{b^2}{a^2}\right]\,,
\label{eq:C6}
\end{equation}
with $\,_{2} F_{1}\left(a,\,b,\,c,\, x\right)$ being the hypergeometric function, $J_{\nu}(x)$ the Bessel function of the first kind and $j_l(x)=\sqrt{\frac{\pi}{2x}}J_{\nu+\frac{1}{2}}(x)$, we can evaluate Eq.~(\ref{eq:C1}) analytically, which yields
\begin{eqnarray}
\chi_{n l}(p) &=& \sum_{k} C_{n l k} 2^{n_{l k}-l}\left(\frac{2 \pi a_{0}}{Z_{l k}}\right)^{3 / 2}\left(\frac{i p a_{0}}{Z_{l k}}\right)^{l} \frac{\Gamma\left(n_{l k}+l+2\right)}{\Gamma(l+\frac{3}{2})\sqrt{\left(2 n_{l k}\right) !}} \nonumber \\
&\times& \,_{2} F_{1}\left[\frac{1}{2}\left(n_{l k}+l+2\right), \frac{1}{2}\left(n_{l k}+l+3\right), l+\frac{3}{2},-\left(\frac{p a_{0}}{Z_{l k}}\right)^{2}\right]\,.
\label{eq:C7}
\end{eqnarray}
We notice that Eq.~(\ref{eq:C7}) has a slightly different expression from Eq.~(C3) in Ref.~\cite{Kopp:2009et}, which leads to a small difference of $\chi_{n l}(p)$ value especially for high $l$. As a crosscheck, we have performed full numerical integration to Eq.~(\ref{eq:C5}) for sample points and found a good agreement with our analytical result.

\section{Modeling of the electron and photonelectron Yields}

We provide additional details to convert the recoiling electron's recoil energy into a specific number of electrons. Our modeling procedure is closely follow Refs.~\cite{Essig:2012yx,Essig:2017kqs}. A primary electron with deposit energy $\Delta E_e= E_R+|E^{nl}_B|$ can produce $n_e$ observable electrons, $n_\gamma$ unobservable scintillation photons and heat. The relevant quantities satisfy following relations
\begin{eqnarray}
E_R&=&(n_\gamma+n_e)W \nonumber \,, \\
n_\gamma&=&N_{\rm ex} + f_R N_i \nonumber\,, \\
n_e&=&(1-f_R)N_i\,.
\end{eqnarray}
Here $W=13.8$ eV is the average energy required to produce a single quanta (photon or electron), $N_{i}$ and $N_{\rm ex}$ are corresponding numbers of ions and excited atoms created by $E_R$ and follow $N_{\rm ex}/N_i\simeq 0.2$~\cite{Doke:2002oab} at energies above a keV. $f_R$ is the fraction of ions that can recombine, and we assume $f_R=0$ at low energy~\cite{Thomas:1987zz}. This then implies that $n_e=N_i$ and $n_\gamma=N_{\rm ex}$, and the fraction of initial quanta observed as electrons is given by~\cite{Sorensen:2011bd}
\begin{equation}
f_e=\frac{n_e}{n_e+n_\gamma}=\frac{1-f_R}{1+N_{\rm ex}/N_i}\simeq 0.83.
\end{equation}

\begin{figure}[ht!]
\centering
\subfigbottomskip=-100pt
\subfigcapskip=-15pt
\subfigure[\,Ultralight mediator, $m_\chi=10$~eV]{ \includegraphics[width=0.4\textwidth]{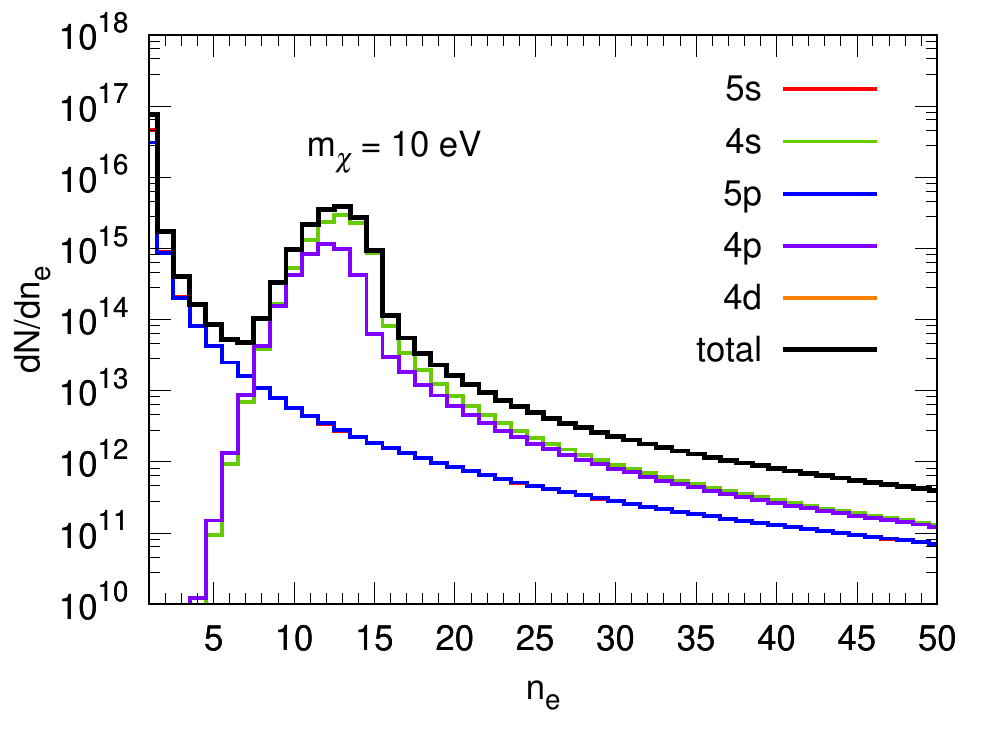}}
\subfigure[\,Ultralight mediator, $m_\chi=1$~keV]{ \includegraphics[width=0.4\textwidth]{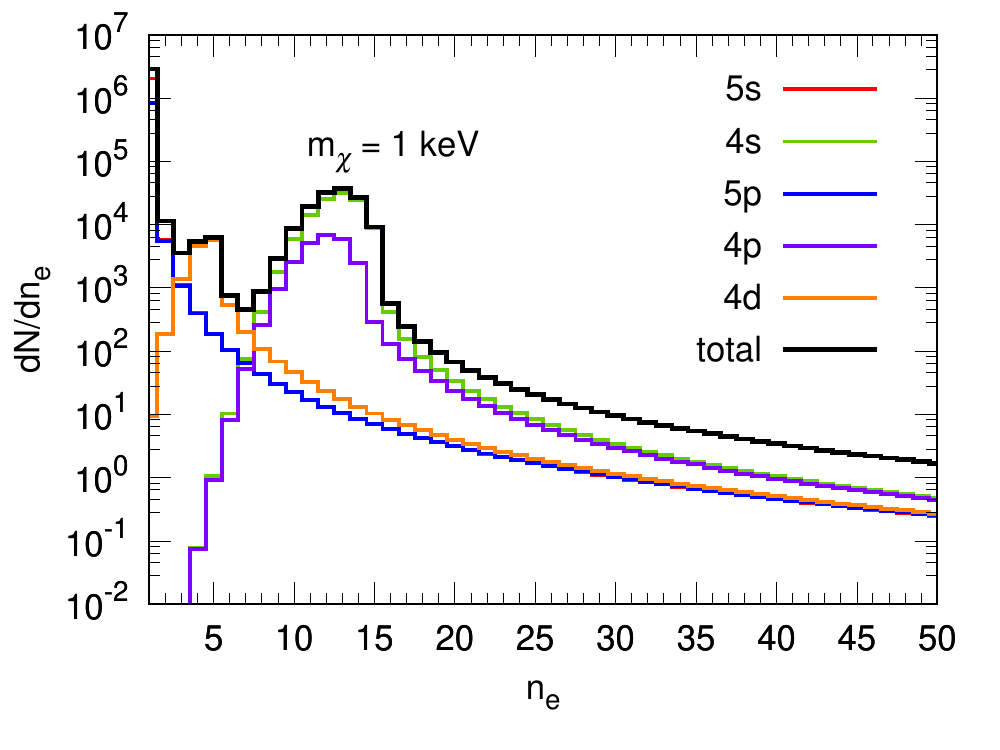}}
\subfigure[\,Heavy mediator, $m_\chi=10$~eV]{ \includegraphics[width=0.4\textwidth]{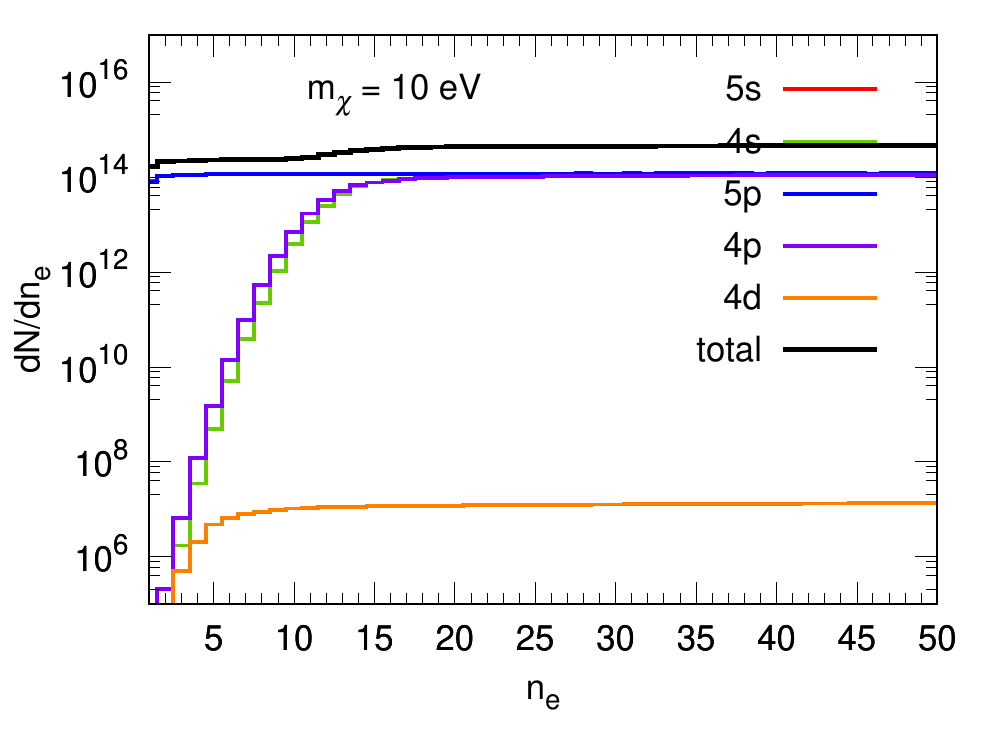}}
\subfigure[\,Heavy mediator, $m_\chi=1$~keV]{ \includegraphics[width=0.4\textwidth]{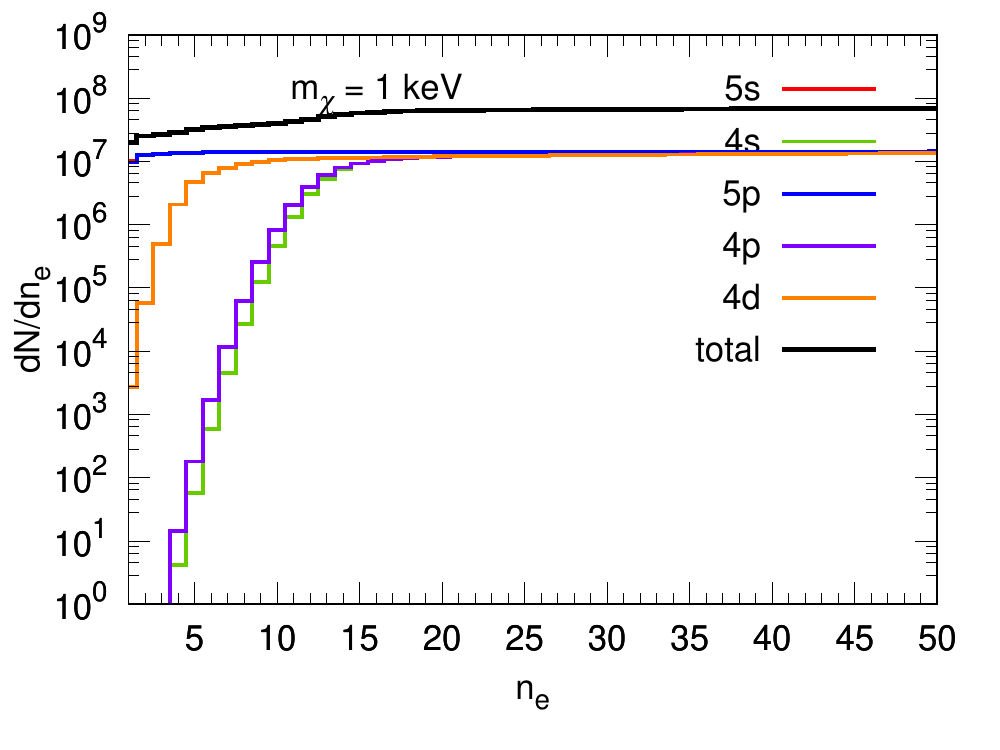}}
\caption{Differential rate $dN/dn_e$ versus number of electrons $n_e$ for $m_\chi=10$ eV and 1 keV, where top (bottom) panel corresponding to ultralight (heavy) mediator cases. The colored lines present the contributions from various xenon shells and the black lines show total contributions.}
\label{fig:nesp-full}
\end{figure}

Furthermore, we assume that the photons associated with the de-excitation of the next-to-outer shells can photoionize to create an additional $n^{(2)}_Q$ quanta, which is listed in Table~\ref{table:n2quanta} for full Xenon electron shells. While in the calculation, we only consider contributions from $(5p^6,5s^2,4d^{10},4p^6,4s^2)$ shells. The total number of electrons is given by $n_e=n^{(1)}_e+n^{(2)}_e$, where $n^{(1)}_e$ is the primary electron and $n^{(2)}$ are the secondary electrons produced. $n^{(1)}$ equals to 0 and 1 with probability $f_R$ and $1-f_R$ respectively, and $n^{(2)}_e$ follows a binomial distribution with $n^{(1)}_Q+n^{(2)}_Q$ trials and success probability $f_e$. As an example, in Fig.~\ref{fig:nesp-full} we plot differential rate $dN/dn_e$ as a function of number of electrons $n_e$ for both ultralight and heavy mediator cases.

\begin{table}[ht!]
\begin{center}
\begin{tabular}{|c||c|c|c|c|c||c|c|c|c|c|c|}
\hline
 Shell & $5p^6$ & $5s^2$ &$4d^{10}$& $4p^6$& $4s^2$ & $3d^{10}$ & $3p^6$ & $3s^2$ & $2p^6$ & $2s^2$ & $1s^2$\\ \hline
 $|E^{nl}_B|$ [eV] &12.4 & 25.7 & 75.6 & 163.5 &  213.8 & 710.7 & 958.4 & 1093.2 & 4837.7& 5152.2 & 33317.6 \\ \hline
 $n^{(2)}_Q$ & 0 & 0 & 4 & 6-9 & 3-14 & 36-50 & 17-68&  9-78 & 271-349 & 22-372 & 2040-2431\\ \hline
\end{tabular}
\caption{Binding energy and number of additional quanta for full Xenon electron shells.
\vspace{-6mm}}
\label{table:n2quanta}
\end{center}
\end{table}%

%% file: draft.bbl
\begin{thebibliography}{49}%
\makeatletter
\providecommand \@ifxundefined [1]{%
 \@ifx{#1\undefined}
}%
\providecommand \@ifnum [1]{%
 \ifnum #1\expandafter \@firstoftwo
 \else \expandafter \@secondoftwo
 \fi
}%
\providecommand \@ifx [1]{%
 \ifx #1\expandafter \@firstoftwo
 \else \expandafter \@secondoftwo
 \fi
}%
\providecommand \natexlab [1]{#1}%
\providecommand \enquote  [1]{``#1''}%
\providecommand \bibnamefont  [1]{#1}%
\providecommand \bibfnamefont [1]{#1}%
\providecommand \citenamefont [1]{#1}%
\providecommand \href@noop [0]{\@secondoftwo}%
\providecommand \href [0]{\begingroup \@sanitize@url \@href}%
\providecommand \@href[1]{\@@startlink{#1}\@@href}%
\providecommand \@@href[1]{\endgroup#1\@@endlink}%
\providecommand \@sanitize@url [0]{\catcode `\\12\catcode `\$12\catcode
  `\&12\catcode `\#12\catcode `\^12\catcode `\_12\catcode `\%12\relax}%
\providecommand \@@startlink[1]{}%
\providecommand \@@endlink[0]{}%
\providecommand \url  [0]{\begingroup\@sanitize@url \@url }%
\providecommand \@url [1]{\endgroup\@href {#1}{\urlprefix }}%
\providecommand \urlprefix  [0]{URL }%
\providecommand \Eprint [0]{\href }%
\providecommand \doibase [0]{http://dx.doi.org/}%
\providecommand \selectlanguage [0]{\@gobble}%
\providecommand \bibinfo  [0]{\@secondoftwo}%
\providecommand \bibfield  [0]{\@secondoftwo}%
\providecommand \translation [1]{[#1]}%
\providecommand \BibitemOpen [0]{}%
\providecommand \bibitemStop [0]{}%
\providecommand \bibitemNoStop [0]{.\EOS\space}%
\providecommand \EOS [0]{\spacefactor3000\relax}%
\providecommand \BibitemShut  [1]{\csname bibitem#1\endcsname}%
\let\auto@bib@innerbib\@empty
\bibitem [{\citenamefont {Hall}\ \emph {et~al.}(2010)\citenamefont {Hall},
  \citenamefont {Jedamzik}, \citenamefont {March-Russell},\ and\ \citenamefont
  {West}}]{Hall:2009bx}%
  \BibitemOpen
  \bibfield  {author} {\bibinfo {author} {\bibfnamefont {L.~J.}\ \bibnamefont
  {Hall}}, \bibinfo {author} {\bibfnamefont {K.}~\bibnamefont {Jedamzik}},
  \bibinfo {author} {\bibfnamefont {J.}~\bibnamefont {March-Russell}}, \ and\
  \bibinfo {author} {\bibfnamefont {S.~M.}\ \bibnamefont {West}},\ }\href
  {\doibase 10.1007/JHEP03(2010)080} {\bibfield  {journal} {\bibinfo  {journal}
  {JHEP}\ }\textbf {\bibinfo {volume} {03}},\ \bibinfo {pages} {080} (\bibinfo
  {year} {2010})},\ \Eprint {http://arxiv.org/abs/0911.1120} {arXiv:0911.1120
  [hep-ph]} \BibitemShut {NoStop}%
\bibitem [{\citenamefont {Chu}\ \emph {et~al.}(2012)\citenamefont {Chu},
  \citenamefont {Hambye},\ and\ \citenamefont {Tytgat}}]{Chu:2011be}%
  \BibitemOpen
  \bibfield  {author} {\bibinfo {author} {\bibfnamefont {X.}~\bibnamefont
  {Chu}}, \bibinfo {author} {\bibfnamefont {T.}~\bibnamefont {Hambye}}, \ and\
  \bibinfo {author} {\bibfnamefont {M.~H.~G.}\ \bibnamefont {Tytgat}},\ }\href
  {\doibase 10.1088/1475-7516/2012/05/034} {\bibfield  {journal} {\bibinfo
  {journal} {JCAP}\ }\textbf {\bibinfo {volume} {1205}},\ \bibinfo {pages}
  {034} (\bibinfo {year} {2012})},\ \Eprint {http://arxiv.org/abs/1112.0493}
  {arXiv:1112.0493 [hep-ph]} \BibitemShut {NoStop}%
\bibitem [{\citenamefont {Essig}\ \emph
  {et~al.}(2012{\natexlab{a}})\citenamefont {Essig}, \citenamefont {Mardon},\
  and\ \citenamefont {Volansky}}]{Essig:2011nj}%
  \BibitemOpen
  \bibfield  {author} {\bibinfo {author} {\bibfnamefont {R.}~\bibnamefont
  {Essig}}, \bibinfo {author} {\bibfnamefont {J.}~\bibnamefont {Mardon}}, \
  and\ \bibinfo {author} {\bibfnamefont {T.}~\bibnamefont {Volansky}},\ }\href
  {\doibase 10.1103/PhysRevD.85.076007} {\bibfield  {journal} {\bibinfo
  {journal} {Phys. Rev.}\ }\textbf {\bibinfo {volume} {D85}},\ \bibinfo {pages}
  {076007} (\bibinfo {year} {2012}{\natexlab{a}})},\ \Eprint
  {http://arxiv.org/abs/1108.5383} {arXiv:1108.5383 [hep-ph]} \BibitemShut
  {NoStop}%
\bibitem [{\citenamefont {Knapen}\ \emph {et~al.}(2017)\citenamefont {Knapen},
  \citenamefont {Lin},\ and\ \citenamefont {Zurek}}]{Knapen:2017xzo}%
  \BibitemOpen
  \bibfield  {author} {\bibinfo {author} {\bibfnamefont {S.}~\bibnamefont
  {Knapen}}, \bibinfo {author} {\bibfnamefont {T.}~\bibnamefont {Lin}}, \ and\
  \bibinfo {author} {\bibfnamefont {K.~M.}\ \bibnamefont {Zurek}},\ }\href
  {\doibase 10.1103/PhysRevD.96.115021} {\bibfield  {journal} {\bibinfo
  {journal} {Phys. Rev.}\ }\textbf {\bibinfo {volume} {D96}},\ \bibinfo {pages}
  {115021} (\bibinfo {year} {2017})},\ \Eprint
  {http://arxiv.org/abs/1709.07882} {arXiv:1709.07882 [hep-ph]} \BibitemShut
  {NoStop}%
\bibitem [{\citenamefont {Bernal}\ \emph {et~al.}(2017)\citenamefont {Bernal},
  \citenamefont {Heikinheimo}, \citenamefont {Tenkanen}, \citenamefont
  {Tuominen},\ and\ \citenamefont {Vaskonen}}]{Bernal:2017kxu}%
  \BibitemOpen
  \bibfield  {author} {\bibinfo {author} {\bibfnamefont {N.}~\bibnamefont
  {Bernal}}, \bibinfo {author} {\bibfnamefont {M.}~\bibnamefont {Heikinheimo}},
  \bibinfo {author} {\bibfnamefont {T.}~\bibnamefont {Tenkanen}}, \bibinfo
  {author} {\bibfnamefont {K.}~\bibnamefont {Tuominen}}, \ and\ \bibinfo
  {author} {\bibfnamefont {V.}~\bibnamefont {Vaskonen}},\ }\href {\doibase
  10.1142/S0217751X1730023X} {\bibfield  {journal} {\bibinfo  {journal} {Int.
  J. Mod. Phys.}\ }\textbf {\bibinfo {volume} {A32}},\ \bibinfo {pages}
  {1730023} (\bibinfo {year} {2017})},\ \Eprint
  {http://arxiv.org/abs/1706.07442} {arXiv:1706.07442 [hep-ph]} \BibitemShut
  {NoStop}%
\bibitem [{\citenamefont {Boehm}\ \emph {et~al.}(2013)\citenamefont {Boehm},
  \citenamefont {Dolan},\ and\ \citenamefont {McCabe}}]{Boehm:2013jpa}%
  \BibitemOpen
  \bibfield  {author} {\bibinfo {author} {\bibfnamefont {C.}~\bibnamefont
  {Boehm}}, \bibinfo {author} {\bibfnamefont {M.~J.}\ \bibnamefont {Dolan}}, \
  and\ \bibinfo {author} {\bibfnamefont {C.}~\bibnamefont {McCabe}},\ }\href
  {\doibase 10.1088/1475-7516/2013/08/041} {\bibfield  {journal} {\bibinfo
  {journal} {JCAP}\ }\textbf {\bibinfo {volume} {1308}},\ \bibinfo {pages}
  {041} (\bibinfo {year} {2013})},\ \Eprint {http://arxiv.org/abs/1303.6270}
  {arXiv:1303.6270 [hep-ph]} \BibitemShut {NoStop}%
\bibitem [{\citenamefont {Nollett}\ and\ \citenamefont
  {Steigman}(2014)}]{Nollett:2013pwa}%
  \BibitemOpen
  \bibfield  {author} {\bibinfo {author} {\bibfnamefont {K.~M.}\ \bibnamefont
  {Nollett}}\ and\ \bibinfo {author} {\bibfnamefont {G.}~\bibnamefont
  {Steigman}},\ }\href {\doibase 10.1103/PhysRevD.89.083508} {\bibfield
  {journal} {\bibinfo  {journal} {Phys. Rev.}\ }\textbf {\bibinfo {volume}
  {D89}},\ \bibinfo {pages} {083508} (\bibinfo {year} {2014})},\ \Eprint
  {http://arxiv.org/abs/1312.5725} {arXiv:1312.5725 [astro-ph.CO]} \BibitemShut
  {NoStop}%
\bibitem [{\citenamefont {Cao}\ \emph {et~al.}(2019)\citenamefont {Cao},
  \citenamefont {Gong}, \citenamefont {Xie},\ and\ \citenamefont
  {Zhang}}]{Cao:2018nbr}%
  \BibitemOpen
  \bibfield  {author} {\bibinfo {author} {\bibfnamefont {Q.-H.}\ \bibnamefont
  {Cao}}, \bibinfo {author} {\bibfnamefont {T.}~\bibnamefont {Gong}}, \bibinfo
  {author} {\bibfnamefont {K.-P.}\ \bibnamefont {Xie}}, \ and\ \bibinfo
  {author} {\bibfnamefont {Z.}~\bibnamefont {Zhang}},\ }\href {\doibase
  10.1007/s11433-018-9322-7} {\bibfield  {journal} {\bibinfo  {journal} {Sci.
  China Phys. Mech. Astron.}\ }\textbf {\bibinfo {volume} {62}},\ \bibinfo
  {pages} {981011} (\bibinfo {year} {2019})},\ \Eprint
  {http://arxiv.org/abs/1810.07658} {arXiv:1810.07658 [hep-ph]} \BibitemShut
  {NoStop}%
\bibitem [{\citenamefont {Dvorkin}\ \emph {et~al.}(2019)\citenamefont
  {Dvorkin}, \citenamefont {Lin},\ and\ \citenamefont
  {Schutz}}]{Dvorkin:2019zdi}%
  \BibitemOpen
  \bibfield  {author} {\bibinfo {author} {\bibfnamefont {C.}~\bibnamefont
  {Dvorkin}}, \bibinfo {author} {\bibfnamefont {T.}~\bibnamefont {Lin}}, \ and\
  \bibinfo {author} {\bibfnamefont {K.}~\bibnamefont {Schutz}},\ }\href
  {\doibase 10.1103/PhysRevD.99.115009} {\bibfield  {journal} {\bibinfo
  {journal} {Phys. Rev.}\ }\textbf {\bibinfo {volume} {D99}},\ \bibinfo {pages}
  {115009} (\bibinfo {year} {2019})},\ \Eprint
  {http://arxiv.org/abs/1902.08623} {arXiv:1902.08623 [hep-ph]} \BibitemShut
  {NoStop}%
\bibitem [{\citenamefont {Essig}\ \emph
  {et~al.}(2012{\natexlab{b}})\citenamefont {Essig}, \citenamefont
  {Manalaysay}, \citenamefont {Mardon}, \citenamefont {Sorensen},\ and\
  \citenamefont {Volansky}}]{Essig:2012yx}%
  \BibitemOpen
  \bibfield  {author} {\bibinfo {author} {\bibfnamefont {R.}~\bibnamefont
  {Essig}}, \bibinfo {author} {\bibfnamefont {A.}~\bibnamefont {Manalaysay}},
  \bibinfo {author} {\bibfnamefont {J.}~\bibnamefont {Mardon}}, \bibinfo
  {author} {\bibfnamefont {P.}~\bibnamefont {Sorensen}}, \ and\ \bibinfo
  {author} {\bibfnamefont {T.}~\bibnamefont {Volansky}},\ }\href {\doibase
  10.1103/PhysRevLett.109.021301} {\bibfield  {journal} {\bibinfo  {journal}
  {Phys. Rev. Lett.}\ }\textbf {\bibinfo {volume} {109}},\ \bibinfo {pages}
  {021301} (\bibinfo {year} {2012}{\natexlab{b}})},\ \Eprint
  {http://arxiv.org/abs/1206.2644} {arXiv:1206.2644 [astro-ph.CO]} \BibitemShut
  {NoStop}%
\bibitem [{\citenamefont {Essig}\ \emph {et~al.}(2017)\citenamefont {Essig},
  \citenamefont {Volansky},\ and\ \citenamefont {Yu}}]{Essig:2017kqs}%
  \BibitemOpen
  \bibfield  {author} {\bibinfo {author} {\bibfnamefont {R.}~\bibnamefont
  {Essig}}, \bibinfo {author} {\bibfnamefont {T.}~\bibnamefont {Volansky}}, \
  and\ \bibinfo {author} {\bibfnamefont {T.-T.}\ \bibnamefont {Yu}},\ }\href
  {\doibase 10.1103/PhysRevD.96.043017} {\bibfield  {journal} {\bibinfo
  {journal} {Phys. Rev.}\ }\textbf {\bibinfo {volume} {D96}},\ \bibinfo {pages}
  {043017} (\bibinfo {year} {2017})},\ \Eprint
  {http://arxiv.org/abs/1703.00910} {arXiv:1703.00910 [hep-ph]} \BibitemShut
  {NoStop}%
\bibitem [{\citenamefont {An}\ \emph {et~al.}(2018)\citenamefont {An},
  \citenamefont {Pospelov}, \citenamefont {Pradler},\ and\ \citenamefont
  {Ritz}}]{An:2017ojc}%
  \BibitemOpen
  \bibfield  {author} {\bibinfo {author} {\bibfnamefont {H.}~\bibnamefont
  {An}}, \bibinfo {author} {\bibfnamefont {M.}~\bibnamefont {Pospelov}},
  \bibinfo {author} {\bibfnamefont {J.}~\bibnamefont {Pradler}}, \ and\
  \bibinfo {author} {\bibfnamefont {A.}~\bibnamefont {Ritz}},\ }\href {\doibase
  10.1103/PhysRevLett.120.141801, 10.1103/PhysRevLett.121.259903} {\bibfield
  {journal} {\bibinfo  {journal} {Phys. Rev. Lett.}\ }\textbf {\bibinfo
  {volume} {120}},\ \bibinfo {pages} {141801} (\bibinfo {year} {2018})},\
  \bibinfo {note} {[Erratum: Phys. Rev. Lett.121,no.25,259903(2018)]},\ \Eprint
  {http://arxiv.org/abs/1708.03642} {arXiv:1708.03642 [hep-ph]} \BibitemShut
  {NoStop}%
\bibitem [{\citenamefont {Bringmann}\ and\ \citenamefont
  {Pospelov}(2019)}]{Bringmann:2018cvk}%
  \BibitemOpen
  \bibfield  {author} {\bibinfo {author} {\bibfnamefont {T.}~\bibnamefont
  {Bringmann}}\ and\ \bibinfo {author} {\bibfnamefont {M.}~\bibnamefont
  {Pospelov}},\ }\href {\doibase 10.1103/PhysRevLett.122.171801} {\bibfield
  {journal} {\bibinfo  {journal} {Phys. Rev. Lett.}\ }\textbf {\bibinfo
  {volume} {122}},\ \bibinfo {pages} {171801} (\bibinfo {year} {2019})},\
  \Eprint {http://arxiv.org/abs/1810.10543} {arXiv:1810.10543 [hep-ph]}
  \BibitemShut {NoStop}%
\bibitem [{\citenamefont {Cappiello}\ \emph {et~al.}(2019)\citenamefont
  {Cappiello}, \citenamefont {Ng},\ and\ \citenamefont
  {Beacom}}]{Cappiello:2018hsu}%
  \BibitemOpen
  \bibfield  {author} {\bibinfo {author} {\bibfnamefont {C.~V.}\ \bibnamefont
  {Cappiello}}, \bibinfo {author} {\bibfnamefont {K.~C.~Y.}\ \bibnamefont
  {Ng}}, \ and\ \bibinfo {author} {\bibfnamefont {J.~F.}\ \bibnamefont
  {Beacom}},\ }\href {\doibase 10.1103/PhysRevD.99.063004} {\bibfield
  {journal} {\bibinfo  {journal} {Phys. Rev.}\ }\textbf {\bibinfo {volume}
  {D99}},\ \bibinfo {pages} {063004} (\bibinfo {year} {2019})},\ \Eprint
  {http://arxiv.org/abs/1810.07705} {arXiv:1810.07705 [hep-ph]} \BibitemShut
  {NoStop}%
\bibitem [{\citenamefont {Ema}\ \emph {et~al.}(2019)\citenamefont {Ema},
  \citenamefont {Sala},\ and\ \citenamefont {Sato}}]{Ema:2018bih}%
  \BibitemOpen
  \bibfield  {author} {\bibinfo {author} {\bibfnamefont {Y.}~\bibnamefont
  {Ema}}, \bibinfo {author} {\bibfnamefont {F.}~\bibnamefont {Sala}}, \ and\
  \bibinfo {author} {\bibfnamefont {R.}~\bibnamefont {Sato}},\ }\href {\doibase
  10.1103/PhysRevLett.122.181802} {\bibfield  {journal} {\bibinfo  {journal}
  {Phys. Rev. Lett.}\ }\textbf {\bibinfo {volume} {122}},\ \bibinfo {pages}
  {181802} (\bibinfo {year} {2019})},\ \Eprint
  {http://arxiv.org/abs/1811.00520} {arXiv:1811.00520 [hep-ph]} \BibitemShut
  {NoStop}%
\bibitem [{\citenamefont {Alvey}\ \emph {et~al.}(2020)\citenamefont {Alvey},
  \citenamefont {Campos}, \citenamefont {Fairbairn},\ and\ \citenamefont
  {You}}]{Alvey:2019zaa}%
  \BibitemOpen
  \bibfield  {author} {\bibinfo {author} {\bibfnamefont {J.}~\bibnamefont
  {Alvey}}, \bibinfo {author} {\bibfnamefont {M.}~\bibnamefont {Campos}},
  \bibinfo {author} {\bibfnamefont {M.}~\bibnamefont {Fairbairn}}, \ and\
  \bibinfo {author} {\bibfnamefont {T.}~\bibnamefont {You}},\ }\href {\doibase
  10.1103/PhysRevLett.123.261802} {\bibfield  {journal} {\bibinfo  {journal}
  {Phys. Rev. Lett.}\ }\textbf {\bibinfo {volume} {123}},\ \bibinfo {pages}
  {261802} (\bibinfo {year} {2020})},\ \bibinfo {note} {[Phys. Rev.
  Lett.123,261802(2019)]},\ \Eprint {http://arxiv.org/abs/1905.05776}
  {arXiv:1905.05776 [hep-ph]} \BibitemShut {NoStop}%
\bibitem [{\citenamefont {Cappiello}\ and\ \citenamefont
  {Beacom}(2019)}]{Cappiello:2019qsw}%
  \BibitemOpen
  \bibfield  {author} {\bibinfo {author} {\bibfnamefont {C.}~\bibnamefont
  {Cappiello}}\ and\ \bibinfo {author} {\bibfnamefont {J.~F.}\ \bibnamefont
  {Beacom}},\ }\href {\doibase 10.1103/PhysRevD.100.103011} {\bibfield
  {journal} {\bibinfo  {journal} {Phys. Rev.}\ }\textbf {\bibinfo {volume}
  {D100}},\ \bibinfo {pages} {103011} (\bibinfo {year} {2019})},\ \Eprint
  {http://arxiv.org/abs/1906.11283} {arXiv:1906.11283 [hep-ph]} \BibitemShut
  {NoStop}%
\bibitem [{\citenamefont {Dent}\ \emph {et~al.}(2019)\citenamefont {Dent},
  \citenamefont {Dutta}, \citenamefont {Newstead},\ and\ \citenamefont
  {Shoemaker}}]{Dent:2019krz}%
  \BibitemOpen
  \bibfield  {author} {\bibinfo {author} {\bibfnamefont {J.~B.}\ \bibnamefont
  {Dent}}, \bibinfo {author} {\bibfnamefont {B.}~\bibnamefont {Dutta}},
  \bibinfo {author} {\bibfnamefont {J.~L.}\ \bibnamefont {Newstead}}, \ and\
  \bibinfo {author} {\bibfnamefont {I.~M.}\ \bibnamefont {Shoemaker}},\
  }\href@noop {} {\  (\bibinfo {year} {2019})},\ \Eprint
  {http://arxiv.org/abs/1907.03782} {arXiv:1907.03782 [hep-ph]} \BibitemShut
  {NoStop}%
\bibitem [{\citenamefont {Krnjaic}\ and\ \citenamefont
  {McDermott}(2019)}]{Krnjaic:2019dzc}%
  \BibitemOpen
  \bibfield  {author} {\bibinfo {author} {\bibfnamefont {G.}~\bibnamefont
  {Krnjaic}}\ and\ \bibinfo {author} {\bibfnamefont {S.~D.}\ \bibnamefont
  {McDermott}},\ }\href@noop {} {\  (\bibinfo {year} {2019})},\ \Eprint
  {http://arxiv.org/abs/1908.00007} {arXiv:1908.00007 [hep-ph]} \BibitemShut
  {NoStop}%
\bibitem [{\citenamefont {Bondarenko}\ \emph {et~al.}(2019)\citenamefont
  {Bondarenko}, \citenamefont {Boyarsky}, \citenamefont {Bringmann},
  \citenamefont {Hufnagel}, \citenamefont {Schmidt-Hoberg},\ and\ \citenamefont
  {Sokolenko}}]{Bondarenko:2019vrb}%
  \BibitemOpen
  \bibfield  {author} {\bibinfo {author} {\bibfnamefont {K.}~\bibnamefont
  {Bondarenko}}, \bibinfo {author} {\bibfnamefont {A.}~\bibnamefont
  {Boyarsky}}, \bibinfo {author} {\bibfnamefont {T.}~\bibnamefont {Bringmann}},
  \bibinfo {author} {\bibfnamefont {M.}~\bibnamefont {Hufnagel}}, \bibinfo
  {author} {\bibfnamefont {K.}~\bibnamefont {Schmidt-Hoberg}}, \ and\ \bibinfo
  {author} {\bibfnamefont {A.}~\bibnamefont {Sokolenko}},\ }\href@noop {} {\
  (\bibinfo {year} {2019})},\ \Eprint {http://arxiv.org/abs/1909.08632}
  {arXiv:1909.08632 [hep-ph]} \BibitemShut {NoStop}%
\bibitem [{\citenamefont {Berger}\ \emph {et~al.}(2019)\citenamefont {Berger},
  \citenamefont {Cui}, \citenamefont {Graham}, \citenamefont {Necib},
  \citenamefont {Petrillo}, \citenamefont {Stocks}, \citenamefont {Tsai},\ and\
  \citenamefont {Zhao}}]{Berger:2019ttc}%
  \BibitemOpen
  \bibfield  {author} {\bibinfo {author} {\bibfnamefont {J.}~\bibnamefont
  {Berger}}, \bibinfo {author} {\bibfnamefont {Y.}~\bibnamefont {Cui}},
  \bibinfo {author} {\bibfnamefont {M.}~\bibnamefont {Graham}}, \bibinfo
  {author} {\bibfnamefont {L.}~\bibnamefont {Necib}}, \bibinfo {author}
  {\bibfnamefont {G.}~\bibnamefont {Petrillo}}, \bibinfo {author}
  {\bibfnamefont {D.}~\bibnamefont {Stocks}}, \bibinfo {author} {\bibfnamefont
  {Y.-T.}\ \bibnamefont {Tsai}}, \ and\ \bibinfo {author} {\bibfnamefont
  {Y.}~\bibnamefont {Zhao}},\ }\href@noop {} {\  (\bibinfo {year} {2019})},\
  \Eprint {http://arxiv.org/abs/1912.05558} {arXiv:1912.05558 [hep-ph]}
  \BibitemShut {NoStop}%
\bibitem [{\citenamefont {Wang}\ \emph {et~al.}(2019)\citenamefont {Wang},
  \citenamefont {Wu}, \citenamefont {Yang}, \citenamefont {Zhou},\ and\
  \citenamefont {Zhu}}]{Wang:2019jtk}%
  \BibitemOpen
  \bibfield  {author} {\bibinfo {author} {\bibfnamefont {W.}~\bibnamefont
  {Wang}}, \bibinfo {author} {\bibfnamefont {L.}~\bibnamefont {Wu}}, \bibinfo
  {author} {\bibfnamefont {J.~M.}\ \bibnamefont {Yang}}, \bibinfo {author}
  {\bibfnamefont {H.}~\bibnamefont {Zhou}}, \ and\ \bibinfo {author}
  {\bibfnamefont {B.}~\bibnamefont {Zhu}},\ }\href@noop {} {\  (\bibinfo {year}
  {2019})},\ \Eprint {http://arxiv.org/abs/1912.09904} {arXiv:1912.09904
  [hep-ph]} \BibitemShut {NoStop}%
\bibitem [{\citenamefont {Goldstein}\ \emph {et~al.}(2001)\citenamefont
  {Goldstein}, \citenamefont {Poole},\ and\ \citenamefont
  {Safko}}]{goldstein:2001}%
  \BibitemOpen
  \bibfield  {author} {\bibinfo {author} {\bibfnamefont {H.}~\bibnamefont
  {Goldstein}}, \bibinfo {author} {\bibfnamefont {C.~P.}\ \bibnamefont
  {Poole}}, \ and\ \bibinfo {author} {\bibfnamefont {J.~L.}\ \bibnamefont
  {Safko}},\ }\href@noop {} {\emph {\bibinfo {title} {Classical Mechanics}}},\
  \bibinfo {edition} {3rd}\ ed.\ (\bibinfo  {publisher} {Addison-Wesley},\
  \bibinfo {year} {2001})\BibitemShut {NoStop}%
\bibitem [{\citenamefont {Navarro}\ \emph {et~al.}(1996)\citenamefont
  {Navarro}, \citenamefont {Frenk},\ and\ \citenamefont
  {White}}]{Navarro:1995iw}%
  \BibitemOpen
  \bibfield  {author} {\bibinfo {author} {\bibfnamefont {J.~F.}\ \bibnamefont
  {Navarro}}, \bibinfo {author} {\bibfnamefont {C.~S.}\ \bibnamefont {Frenk}},
  \ and\ \bibinfo {author} {\bibfnamefont {S.~D.~M.}\ \bibnamefont {White}},\
  }\href {\doibase 10.1086/177173} {\bibfield  {journal} {\bibinfo  {journal}
  {Astrophys. J.}\ }\textbf {\bibinfo {volume} {462}},\ \bibinfo {pages} {563}
  (\bibinfo {year} {1996})},\ \Eprint {http://arxiv.org/abs/astro-ph/9508025}
  {arXiv:astro-ph/9508025 [astro-ph]} \BibitemShut {NoStop}%
\bibitem [{\citenamefont {Navarro}\ \emph {et~al.}(1997)\citenamefont
  {Navarro}, \citenamefont {Frenk},\ and\ \citenamefont
  {White}}]{Navarro:1996gj}%
  \BibitemOpen
  \bibfield  {author} {\bibinfo {author} {\bibfnamefont {J.~F.}\ \bibnamefont
  {Navarro}}, \bibinfo {author} {\bibfnamefont {C.~S.}\ \bibnamefont {Frenk}},
  \ and\ \bibinfo {author} {\bibfnamefont {S.~D.~M.}\ \bibnamefont {White}},\
  }\href {\doibase 10.1086/304888} {\bibfield  {journal} {\bibinfo  {journal}
  {Astrophys. J.}\ }\textbf {\bibinfo {volume} {490}},\ \bibinfo {pages} {493}
  (\bibinfo {year} {1997})},\ \Eprint {http://arxiv.org/abs/astro-ph/9611107}
  {arXiv:astro-ph/9611107 [astro-ph]} \BibitemShut {NoStop}%
\bibitem [{\citenamefont {Holdom}(1986)}]{Holdom:1985ag}%
  \BibitemOpen
  \bibfield  {author} {\bibinfo {author} {\bibfnamefont {B.}~\bibnamefont
  {Holdom}},\ }\href {\doibase 10.1016/0370-2693(86)91377-8} {\bibfield
  {journal} {\bibinfo  {journal} {Phys. Lett.}\ }\textbf {\bibinfo {volume}
  {166B}},\ \bibinfo {pages} {196} (\bibinfo {year} {1986})}\BibitemShut
  {NoStop}%
\bibitem [{\citenamefont {Essig}\ \emph {et~al.}(2016)\citenamefont {Essig},
  \citenamefont {Fernandez-Serra}, \citenamefont {Mardon}, \citenamefont
  {Soto}, \citenamefont {Volansky},\ and\ \citenamefont {Yu}}]{Essig:2015cda}%
  \BibitemOpen
  \bibfield  {author} {\bibinfo {author} {\bibfnamefont {R.}~\bibnamefont
  {Essig}}, \bibinfo {author} {\bibfnamefont {M.}~\bibnamefont
  {Fernandez-Serra}}, \bibinfo {author} {\bibfnamefont {J.}~\bibnamefont
  {Mardon}}, \bibinfo {author} {\bibfnamefont {A.}~\bibnamefont {Soto}},
  \bibinfo {author} {\bibfnamefont {T.}~\bibnamefont {Volansky}}, \ and\
  \bibinfo {author} {\bibfnamefont {T.-T.}\ \bibnamefont {Yu}},\ }\href
  {\doibase 10.1007/JHEP05(2016)046} {\bibfield  {journal} {\bibinfo  {journal}
  {JHEP}\ }\textbf {\bibinfo {volume} {05}},\ \bibinfo {pages} {046} (\bibinfo
  {year} {2016})},\ \Eprint {http://arxiv.org/abs/1509.01598} {arXiv:1509.01598
  [hep-ph]} \BibitemShut {NoStop}%
\bibitem [{\citenamefont {Bunge}\ \emph {et~al.}(1993)\citenamefont {Bunge},
  \citenamefont {Barrientos},\ and\ \citenamefont {Bunge}}]{Bunge:1993jsz}%
  \BibitemOpen
  \bibfield  {author} {\bibinfo {author} {\bibfnamefont {C.~F.}\ \bibnamefont
  {Bunge}}, \bibinfo {author} {\bibfnamefont {J.~A.}\ \bibnamefont
  {Barrientos}}, \ and\ \bibinfo {author} {\bibfnamefont {A.~V.}\ \bibnamefont
  {Bunge}},\ }\href {\doibase 10.1006/adnd.1993.1003} {\bibfield  {journal}
  {\bibinfo  {journal} {Atom. Data Nucl. Data Tabl.}\ }\textbf {\bibinfo
  {volume} {53}},\ \bibinfo {pages} {113} (\bibinfo {year} {1993})}\BibitemShut
  {NoStop}%
\bibitem [{\citenamefont {Aprile}\ \emph {et~al.}(2016)\citenamefont {Aprile}
  \emph {et~al.}}]{Aprile:2016wwo}%
  \BibitemOpen
  \bibfield  {author} {\bibinfo {author} {\bibfnamefont {E.}~\bibnamefont
  {Aprile}} \emph {et~al.} (\bibinfo {collaboration} {XENON}),\ }\href
  {\doibase 10.1103/PhysRevD.94.092001, 10.1103/PhysRevD.95.059901} {\bibfield
  {journal} {\bibinfo  {journal} {Phys. Rev.}\ }\textbf {\bibinfo {volume}
  {D94}},\ \bibinfo {pages} {092001} (\bibinfo {year} {2016})},\ \bibinfo
  {note} {[Erratum: Phys. Rev.D95,no.5,059901(2017)]},\ \Eprint
  {http://arxiv.org/abs/1605.06262} {arXiv:1605.06262 [astro-ph.CO]}
  \BibitemShut {NoStop}%
\bibitem [{\citenamefont {Aprile}\ \emph
  {et~al.}(2019{\natexlab{a}})\citenamefont {Aprile} \emph
  {et~al.}}]{Aprile:2019xxb}%
  \BibitemOpen
  \bibfield  {author} {\bibinfo {author} {\bibfnamefont {E.}~\bibnamefont
  {Aprile}} \emph {et~al.} (\bibinfo {collaboration} {XENON}),\ }\href
  {\doibase 10.1103/PhysRevLett.123.251801} {\bibfield  {journal} {\bibinfo
  {journal} {Phys. Rev. Lett.}\ }\textbf {\bibinfo {volume} {123}},\ \bibinfo
  {pages} {251801} (\bibinfo {year} {2019}{\natexlab{a}})},\ \Eprint
  {http://arxiv.org/abs/1907.11485} {arXiv:1907.11485 [hep-ex]} \BibitemShut
  {NoStop}%
\bibitem [{\citenamefont {Aprile}\ \emph
  {et~al.}(2019{\natexlab{b}})\citenamefont {Aprile} \emph
  {et~al.}}]{Aprile:2019dme}%
  \BibitemOpen
  \bibfield  {author} {\bibinfo {author} {\bibfnamefont {E.}~\bibnamefont
  {Aprile}} \emph {et~al.} (\bibinfo {collaboration} {XENON}),\ }\href
  {\doibase 10.1103/PhysRevD.99.112009} {\bibfield  {journal} {\bibinfo
  {journal} {Phys. Rev. D}\ }\textbf {\bibinfo {volume} {99}},\ \bibinfo
  {pages} {112009} (\bibinfo {year} {2019}{\natexlab{b}})},\ \Eprint
  {http://arxiv.org/abs/1902.11297} {arXiv:1902.11297 [physics.ins-det]}
  \BibitemShut {NoStop}%
\bibitem [{\citenamefont {Chang}\ \emph {et~al.}(2018)\citenamefont {Chang},
  \citenamefont {Essig},\ and\ \citenamefont {McDermott}}]{Chang:2018rso}%
  \BibitemOpen
  \bibfield  {author} {\bibinfo {author} {\bibfnamefont {J.~H.}\ \bibnamefont
  {Chang}}, \bibinfo {author} {\bibfnamefont {R.}~\bibnamefont {Essig}}, \ and\
  \bibinfo {author} {\bibfnamefont {S.~D.}\ \bibnamefont {McDermott}},\ }\href
  {\doibase 10.1007/JHEP09(2018)051} {\bibfield  {journal} {\bibinfo  {journal}
  {JHEP}\ }\textbf {\bibinfo {volume} {09}},\ \bibinfo {pages} {051} (\bibinfo
  {year} {2018})},\ \Eprint {http://arxiv.org/abs/1803.00993} {arXiv:1803.00993
  [hep-ph]} \BibitemShut {NoStop}%
\bibitem [{\citenamefont {Vogel}\ and\ \citenamefont
  {Redondo}(2014)}]{Vogel:2013raa}%
  \BibitemOpen
  \bibfield  {author} {\bibinfo {author} {\bibfnamefont {H.}~\bibnamefont
  {Vogel}}\ and\ \bibinfo {author} {\bibfnamefont {J.}~\bibnamefont
  {Redondo}},\ }\href {\doibase 10.1088/1475-7516/2014/02/029} {\bibfield
  {journal} {\bibinfo  {journal} {JCAP}\ }\textbf {\bibinfo {volume} {1402}},\
  \bibinfo {pages} {029} (\bibinfo {year} {2014})},\ \Eprint
  {http://arxiv.org/abs/1311.2600} {arXiv:1311.2600 [hep-ph]} \BibitemShut
  {NoStop}%
\bibitem [{\citenamefont {Bays}\ \emph {et~al.}(2012)\citenamefont {Bays} \emph
  {et~al.}}]{Bays:2011si}%
  \BibitemOpen
  \bibfield  {author} {\bibinfo {author} {\bibfnamefont {K.}~\bibnamefont
  {Bays}} \emph {et~al.} (\bibinfo {collaboration} {Super-Kamiokande}),\ }\href
  {\doibase 10.1103/PhysRevD.85.052007} {\bibfield  {journal} {\bibinfo
  {journal} {Phys. Rev.}\ }\textbf {\bibinfo {volume} {D85}},\ \bibinfo {pages}
  {052007} (\bibinfo {year} {2012})},\ \Eprint {http://arxiv.org/abs/1111.5031}
  {arXiv:1111.5031 [hep-ex]} \BibitemShut {NoStop}%
\bibitem [{\citenamefont {Carlson}\ and\ \citenamefont
  {Profumo}(2015)}]{Carlson:2015daa}%
  \BibitemOpen
  \bibfield  {author} {\bibinfo {author} {\bibfnamefont {E.}~\bibnamefont
  {Carlson}}\ and\ \bibinfo {author} {\bibfnamefont {S.}~\bibnamefont
  {Profumo}},\ }\href {\doibase 10.1103/PhysRevD.92.063003} {\bibfield
  {journal} {\bibinfo  {journal} {Phys. Rev.}\ }\textbf {\bibinfo {volume}
  {D92}},\ \bibinfo {pages} {063003} (\bibinfo {year} {2015})},\ \Eprint
  {http://arxiv.org/abs/1504.04782} {arXiv:1504.04782 [astro-ph.HE]}
  \BibitemShut {NoStop}%
\bibitem [{\citenamefont {Cao}\ \emph {et~al.}()\citenamefont {Cao},
  \citenamefont {Ding},\ and\ \citenamefont {Xiang}}]{cao:2019cr}%
  \BibitemOpen
  \bibfield  {author} {\bibinfo {author} {\bibfnamefont {Q.-H.}\ \bibnamefont
  {Cao}}, \bibinfo {author} {\bibfnamefont {R.}~\bibnamefont {Ding}}, \ and\
  \bibinfo {author} {\bibfnamefont {Q.-F.}\ \bibnamefont {Xiang}},\ }\href@noop
  {} {\ }\Eprint {http://arxiv.org/abs/{in preparation}} {{in preparation}}
  \BibitemShut {NoStop}%
\bibitem [{\citenamefont {Aprile}\ \emph {et~al.}(2020)\citenamefont {Aprile}
  \emph {et~al.}}]{Aprile:2020tmw}%
  \BibitemOpen
  \bibfield  {author} {\bibinfo {author} {\bibfnamefont {E.}~\bibnamefont
  {Aprile}} \emph {et~al.} (\bibinfo {collaboration} {XENON}),\ }\href@noop {}
  {\  (\bibinfo {year} {2020})},\ \Eprint {http://arxiv.org/abs/2006.09721}
  {arXiv:2006.09721 [hep-ex]} \BibitemShut {NoStop}%
\bibitem [{\citenamefont {Cummings}\ \emph {et~al.}(2016)\citenamefont
  {Cummings}, \citenamefont {Stone}, \citenamefont {Heikkila}, \citenamefont
  {Lal}, \citenamefont {Webber}, \citenamefont {Jahannesson}, \citenamefont
  {Moskalenko}, \citenamefont {Orlando},\ and\ \citenamefont
  {Porter}}]{Cummings:2016pdr}%
  \BibitemOpen
  \bibfield  {author} {\bibinfo {author} {\bibfnamefont {A.~C.}\ \bibnamefont
  {Cummings}}, \bibinfo {author} {\bibfnamefont {E.~C.}\ \bibnamefont {Stone}},
  \bibinfo {author} {\bibfnamefont {B.~C.}\ \bibnamefont {Heikkila}}, \bibinfo
  {author} {\bibfnamefont {N.}~\bibnamefont {Lal}}, \bibinfo {author}
  {\bibfnamefont {W.~R.}\ \bibnamefont {Webber}}, \bibinfo {author}
  {\bibfnamefont {G.}~\bibnamefont {Jahannesson}}, \bibinfo {author}
  {\bibfnamefont {I.~V.}\ \bibnamefont {Moskalenko}}, \bibinfo {author}
  {\bibfnamefont {E.}~\bibnamefont {Orlando}}, \ and\ \bibinfo {author}
  {\bibfnamefont {T.~A.}\ \bibnamefont {Porter}},\ }\href {\doibase
  10.3847/0004-637X/831/1/18} {\bibfield  {journal} {\bibinfo  {journal}
  {Astrophys. J.}\ }\textbf {\bibinfo {volume} {831}},\ \bibinfo {pages} {18}
  (\bibinfo {year} {2016})}\BibitemShut {NoStop}%
\bibitem [{\citenamefont {Aguilar}\ \emph {et~al.}(2014)\citenamefont {Aguilar}
  \emph {et~al.}}]{Aguilar:2014mma}%
  \BibitemOpen
  \bibfield  {author} {\bibinfo {author} {\bibfnamefont {M.}~\bibnamefont
  {Aguilar}} \emph {et~al.} (\bibinfo {collaboration} {AMS}),\ }\href {\doibase
  10.1103/PhysRevLett.113.121102} {\bibfield  {journal} {\bibinfo  {journal}
  {Phys. Rev. Lett.}\ }\textbf {\bibinfo {volume} {113}},\ \bibinfo {pages}
  {121102} (\bibinfo {year} {2014})}\BibitemShut {NoStop}%
\bibitem [{\citenamefont {Ambrosi}\ \emph {et~al.}(2017)\citenamefont {Ambrosi}
  \emph {et~al.}}]{Ambrosi:2017wek}%
  \BibitemOpen
  \bibfield  {author} {\bibinfo {author} {\bibfnamefont {G.}~\bibnamefont
  {Ambrosi}} \emph {et~al.} (\bibinfo {collaboration} {DAMPE}),\ }\href
  {\doibase 10.1038/nature24475} {\bibfield  {journal} {\bibinfo  {journal}
  {Nature}\ }\textbf {\bibinfo {volume} {552}},\ \bibinfo {pages} {63}
  (\bibinfo {year} {2017})},\ \Eprint {http://arxiv.org/abs/1711.10981}
  {arXiv:1711.10981 [astro-ph.HE]} \BibitemShut {NoStop}%
\bibitem [{\citenamefont {Strong}\ and\ \citenamefont
  {Moskalenko}(1998)}]{Strong:1998pw}%
  \BibitemOpen
  \bibfield  {author} {\bibinfo {author} {\bibfnamefont {A.~W.}\ \bibnamefont
  {Strong}}\ and\ \bibinfo {author} {\bibfnamefont {I.~V.}\ \bibnamefont
  {Moskalenko}},\ }\href {\doibase 10.1086/306470} {\bibfield  {journal}
  {\bibinfo  {journal} {Astrophys. J.}\ }\textbf {\bibinfo {volume} {509}},\
  \bibinfo {pages} {212} (\bibinfo {year} {1998})},\ \Eprint
  {http://arxiv.org/abs/astro-ph/9807150} {arXiv:astro-ph/9807150 [astro-ph]}
  \BibitemShut {NoStop}%
\bibitem [{\citenamefont {Moskalenko}\ and\ \citenamefont
  {Strong}(1998)}]{Moskalenko:1997gh}%
  \BibitemOpen
  \bibfield  {author} {\bibinfo {author} {\bibfnamefont {I.~V.}\ \bibnamefont
  {Moskalenko}}\ and\ \bibinfo {author} {\bibfnamefont {A.~W.}\ \bibnamefont
  {Strong}},\ }\href {\doibase 10.1086/305152} {\bibfield  {journal} {\bibinfo
  {journal} {Astrophys. J.}\ }\textbf {\bibinfo {volume} {493}},\ \bibinfo
  {pages} {694} (\bibinfo {year} {1998})},\ \Eprint
  {http://arxiv.org/abs/astro-ph/9710124} {arXiv:astro-ph/9710124 [astro-ph]}
  \BibitemShut {NoStop}%
\bibitem [{\citenamefont {Potgieter}\ \emph {et~al.}(2015)\citenamefont
  {Potgieter}, \citenamefont {Vos}, \citenamefont {Munini}, \citenamefont
  {Boezio},\ and\ \citenamefont {Di~Felice}}]{Potgieter:2015jxa}%
  \BibitemOpen
  \bibfield  {author} {\bibinfo {author} {\bibfnamefont {M.~S.}\ \bibnamefont
  {Potgieter}}, \bibinfo {author} {\bibfnamefont {E.~E.}\ \bibnamefont {Vos}},
  \bibinfo {author} {\bibfnamefont {R.}~\bibnamefont {Munini}}, \bibinfo
  {author} {\bibfnamefont {M.}~\bibnamefont {Boezio}}, \ and\ \bibinfo {author}
  {\bibfnamefont {V.}~\bibnamefont {Di~Felice}},\ }\href {\doibase
  10.1088/0004-637X/810/2/141} {\bibfield  {journal} {\bibinfo  {journal}
  {Astrophys. J.}\ }\textbf {\bibinfo {volume} {810}},\ \bibinfo {pages} {141}
  (\bibinfo {year} {2015})}\BibitemShut {NoStop}%
\bibitem [{\citenamefont {Abdollahi}\ \emph {et~al.}(2017)\citenamefont
  {Abdollahi} \emph {et~al.}}]{Abdollahi:2017nat}%
  \BibitemOpen
  \bibfield  {author} {\bibinfo {author} {\bibfnamefont {S.}~\bibnamefont
  {Abdollahi}} \emph {et~al.} (\bibinfo {collaboration} {Fermi-LAT}),\ }\href
  {\doibase 10.1103/PhysRevD.95.082007} {\bibfield  {journal} {\bibinfo
  {journal} {Phys. Rev.}\ }\textbf {\bibinfo {volume} {D95}},\ \bibinfo {pages}
  {082007} (\bibinfo {year} {2017})},\ \Eprint
  {http://arxiv.org/abs/1704.07195} {arXiv:1704.07195 [astro-ph.HE]}
  \BibitemShut {NoStop}%
\bibitem [{\citenamefont {Kopp}\ \emph {et~al.}(2009)\citenamefont {Kopp},
  \citenamefont {Niro}, \citenamefont {Schwetz},\ and\ \citenamefont
  {Zupan}}]{Kopp:2009et}%
  \BibitemOpen
  \bibfield  {author} {\bibinfo {author} {\bibfnamefont {J.}~\bibnamefont
  {Kopp}}, \bibinfo {author} {\bibfnamefont {V.}~\bibnamefont {Niro}}, \bibinfo
  {author} {\bibfnamefont {T.}~\bibnamefont {Schwetz}}, \ and\ \bibinfo
  {author} {\bibfnamefont {J.}~\bibnamefont {Zupan}},\ }\href {\doibase
  10.1103/PhysRevD.80.083502} {\bibfield  {journal} {\bibinfo  {journal} {Phys.
  Rev.}\ }\textbf {\bibinfo {volume} {D80}},\ \bibinfo {pages} {083502}
  (\bibinfo {year} {2009})},\ \Eprint {http://arxiv.org/abs/0907.3159}
  {arXiv:0907.3159 [hep-ph]} \BibitemShut {NoStop}%
\bibitem [{\citenamefont {Wang}\ and\ \citenamefont {Guo}(1989)}]{wang:1989}%
  \BibitemOpen
  \bibfield  {author} {\bibinfo {author} {\bibfnamefont {Z.~X.}\ \bibnamefont
  {Wang}}\ and\ \bibinfo {author} {\bibfnamefont {D.~R.}\ \bibnamefont {Guo}},\
  }\href@noop {} {\emph {\bibinfo {title} {Special Functions}}}\ (\bibinfo
  {publisher} {World Scientific},\ \bibinfo {year} {1989})\BibitemShut
  {NoStop}%
\bibitem [{\citenamefont {Doke}\ \emph {et~al.}(2002)\citenamefont {Doke},
  \citenamefont {Hitachi}, \citenamefont {Kikuchi}, \citenamefont {Masuda},
  \citenamefont {Okada},\ and\ \citenamefont {Shibamura}}]{Doke:2002oab}%
  \BibitemOpen
  \bibfield  {author} {\bibinfo {author} {\bibfnamefont {T.}~\bibnamefont
  {Doke}}, \bibinfo {author} {\bibfnamefont {A.}~\bibnamefont {Hitachi}},
  \bibinfo {author} {\bibfnamefont {J.}~\bibnamefont {Kikuchi}}, \bibinfo
  {author} {\bibfnamefont {K.}~\bibnamefont {Masuda}}, \bibinfo {author}
  {\bibfnamefont {H.}~\bibnamefont {Okada}}, \ and\ \bibinfo {author}
  {\bibfnamefont {E.}~\bibnamefont {Shibamura}},\ }\href {\doibase
  10.1143/JJAP.41.1538} {\bibfield  {journal} {\bibinfo  {journal} {Jap. J.
  Appl. Phys.}\ }\textbf {\bibinfo {volume} {41}},\ \bibinfo {pages} {1538}
  (\bibinfo {year} {2002})}\BibitemShut {NoStop}%
\bibitem [{\citenamefont {Thomas}\ and\ \citenamefont
  {Imel}(1987)}]{Thomas:1987zz}%
  \BibitemOpen
  \bibfield  {author} {\bibinfo {author} {\bibfnamefont {J.}~\bibnamefont
  {Thomas}}\ and\ \bibinfo {author} {\bibfnamefont {D.~A.}\ \bibnamefont
  {Imel}},\ }\href {\doibase 10.1103/PhysRevA.36.614} {\bibfield  {journal}
  {\bibinfo  {journal} {Phys. Rev.}\ }\textbf {\bibinfo {volume} {A36}},\
  \bibinfo {pages} {614} (\bibinfo {year} {1987})}\BibitemShut {NoStop}%
\bibitem [{\citenamefont {Sorensen}\ and\ \citenamefont
  {Dahl}(2011)}]{Sorensen:2011bd}%
  \BibitemOpen
  \bibfield  {author} {\bibinfo {author} {\bibfnamefont {P.}~\bibnamefont
  {Sorensen}}\ and\ \bibinfo {author} {\bibfnamefont {C.~E.}\ \bibnamefont
  {Dahl}},\ }\href {\doibase 10.1103/PhysRevD.83.063501} {\bibfield  {journal}
  {\bibinfo  {journal} {Phys. Rev.}\ }\textbf {\bibinfo {volume} {D83}},\
  \bibinfo {pages} {063501} (\bibinfo {year} {2011})},\ \Eprint
  {http://arxiv.org/abs/1101.6080} {arXiv:1101.6080 [astro-ph.IM]} \BibitemShut
  {NoStop}%
\end{thebibliography}%
